\documentclass[12pt,final,3p,times,authoryear]{elsarticle}
\usepackage[T1]{fontenc} 
\usepackage[utf8]{inputenc}
\usepackage{graphicx}
\usepackage{verbatim}
\usepackage{hyperref}
\usepackage{fancyhdr}
\expandafter\let\csname equation*\endcsname\relax
\expandafter\let\csname endequation*\endcsname\relax
\usepackage{amsfonts,amsmath,amssymb,mathrsfs,mathtools,suffix}
\usepackage{svg}

\usepackage{refcheck}
\norefnames
\nocitenames

\newcommand{\csch}{\mathrm{csch} \,}
\newcommand{\R}{\mathbb{R}}
\newcommand{\Z}{\mathbb{Z}}
\newcommand{\N}{\mathbb{N}}

\makeatletter
\newcommand{\pushright}[1]{\ifmeasuring@#1\else\omit\hfill$\displaystyle#1$\fi\ignorespaces}
\newcommand{\pushleft}[1]{\ifmeasuring@#1\else\omit$\displaystyle#1$\hfill\fi\ignorespaces}
\makeatother

\DeclarePairedDelimiterX\MeijerM[3]{\lparen}{\rparen}%
{\genfrac{}{}{0pt}{}{#1}{#2} \Bigg| \, #3}
\newcommand\pFq[5]{%
   \, _{#1}F_{#2}\left(\genfrac{}{}{0pt}{}{#3}{#4} \Bigg| \, #5\right)}
\newcommand\MeijerG[8][]{%
  G^{\,#2,#3}_{#4,#5}\MeijerM[#1]{#6}{#7}{#8}}
\WithSuffix\newcommand\MeijerG*[7]{%
  G^{\,#1,#2}_{#3,#4}\MeijerM*{#5}{#6}{#7}}
  
\journal{Nuclear Physics B}

\begin{document}
\begin{frontmatter}
    \title{Anomalous diffusion and factor ordering in (1+1)-dimensional Lorentzian quantum gravity\footnote{\textbf{©2023 The MITRE Corporation. ALL RIGHTS RESERVED.}}}
    \author[wit]{E Sanderson}
    \ead{sandersone1@wit.edu}
    
    \author[wit]{R L Maitra}
    \ead{maitrar@wit.edu}
    
    \author[mitre]{AJ Liberatore\footnote{The author's affiliation with The MITRE Corporation is provided for identification purposes only, and is not intended to convey or imply MITRE's concurrence with, or support for, the positions, opinions, or viewpoints expressed by the author.}\textsuperscript{,}}
    \ead{aliberatore@mitre.org}
    
    \affiliation[wit]{organization={Wentworth Institute of Technology},
        addressline={550 Huntington Avenue},
        city={Boston},
        postcode={02115},
        state={MA},
        country={USA}}
    \affiliation[mitre]{organization={The MITRE Corporation},
        addressline={202 Burlington Road},
        city={Bedford},
        postcode={01730},
        state={MA},
        country={USA}}

    \begin{abstract}
        Using properties of diffusion according to a quantum heat kernel constructed as an expectation over classical heat kernels on $S^1$, we probe the non-manifold-like nature of quantized space in a model of (1+1)-dimensional quantum gravity.  By computing the mean squared displacement of a diffusing particle, we find that diffusion is anomalous, behaving similarly to that on a porous substrate, network, or fractal over short distances.  The walk dimension of the path for a particle diffusing in quantized space is calculated to have an infimum of 4, rising to arbitrarily large values depending on a parameter labeling the choice of factor ordering in the quantum Hamiltonian for our model and figuring in the asymptotic behavior of the wavefunction used to construct the quantum heat kernel.  Additionally, we derive an expansion for return probability of a diffusing particle, whose modifications from the classical power-series form depend on the factor-ordering parameter.
    \end{abstract}
    
    \fancyhead{} % clear all footer fields
    % \fancyhead[L,R]{\thepage}
    \fancyhead[C]{\textbf{©2023 The MITRE Corporation. ALL RIGHTS RESERVED.}}
    
    % \maketitle
\end{frontmatter}

\section{Introduction}

Since John Wheeler first coined the phrase ‘spacetime foam’ nearly seventy years ago \citep{wheeler_geons_1955, wheeler_nature_1957}, a plethora of analytical and computational evidence has emerged to support the conclusion that at subatomic scales, spacetime behaves entirely unlike a smooth manifold (\citet{carlip_spacetime_2023}, and references therein).  However, the process of quantization is rich with nuance, and the choices made should influence the described character of quantum geometry.  In particular, the definition of a path integral measure over a set of spacetime geometries, or equivalently the definition of a Wheeler-DeWitt operator on a space of physical wavefunctions, involves choices which determine essential aspects of the quantization.  

To investigate how factor ordering of the canonically quantized Wheeler-DeWitt operator affects predictions for quantum geometry, we use a reduced model of Lorentzian (1+1)-dimensional gravity as in \citet{nakayama_2d_1994}, where the constructions involved are more rigorously understood than in the physical (3+1)-dimensional case.  We find that diffusion in quantized space is non-Gaussian and anomalous:  the mean squared displacement of a particle diffusing through space according to our quantum-averaged heat kernel does not obey the standard scaling $\langle \Delta x^2 \rangle \propto t$ for small diffusion time $t$, but instead is \emph{subdiffusive}, so that if diffusion time is subdivided into shorter intervals, the particle has on average a smaller displacement over the full diffusion time than expected based on linear extrapolation from subintervals.  Consequently the walk dimension of a diffusing particle's path is higher than the classical value of 2 expected from standard diffusion.  Moreover the dependence of the mean squared displacement on diffusion time is dictated by asymptotic boundary behavior of the wavefunction used for quantum averaging of our heat kernel, with a wavefunction's bias toward small universes predicting increasingly anomalous diffusion.  The asymptotic behavior of wavefunctions in turn results from the factor ordering chosen for the quantized Hamiltonian.  Additionally, we compute the spectral dimension of space as indicated by return probability from our quantum heat kernel, and find it equal to the classical value of 1 for all factor orderings considered, implying that the disruptions of space due to quantization are detected by some aspects of the diffusion process, while others remain unperturbed. Though the results of the present paper pertain to (1+1)-dimensional gravity, we anticipate the applicability of a similar method to minisuperspace models of quantum cosmology, although surely necessitating a more computational approach than the fully analytic calculations presented here.

Very recently, much progress has been made to understand the anomalous properties of spacetime in Euclidean 2d quantum gravity ($\gamma$-Liouville gravity from a continuum perspective, or discretized models such as random planar maps).  Although distinguished from our approach by their use of Euclidean-signature path integral quantization, these methods relatedly consider a quantum-averaged fictitious diffusion on Euclideanized spacetime to show that its spectral dimension is 2 independent of discretization scheme or value of the parameter $\gamma$ controlling ``roughness" of spacetime via the formally defined metric $ds^2 = e^{\gamma \varphi} d \hat{s}^2$, where $\varphi$ is a random field.  Computations \citep{barkley_precision_2019} or constraints \citep{ding_fractal_2020} on volume scaling of geodesic balls show that the Hausdorff dimension $d_\gamma$ of quantized spacetime is a monotonically increasing function of $\gamma$ taking values strictly greater than 2, although only known exactly in the ``pure gravity'' case $d_{\sqrt{8/3}}=4$ (so called because of $\gamma$-Liouville gravity's correspondence with a 2d metric coupled to $c=25-6\left( 
 \frac{2}{\gamma} + \frac{\gamma}{2}\right)^2$ scalar fields).  Moreover, diffusion on a class of random planar maps in $\gamma$-Liouville quantum gravity proves to be anomalous, exhibiting subdiffusive behavior \citep{gwynne_anomalous_2020}. These results accord with ours in that both approaches show spectral dimension of quantized space(time) retaining classical values, while indicators depending on scaling behavior of space(time) deviate from classicality.  

In the subsections below, we review the model at hand for (1+1)-dimensional quantum gravity (\S \ref{2dqg}), the heat kernel for diffusion on a circle (\S \ref{diffcirc}), and the relation between anomalous diffusion and walk dimension of the path of a diffusing particle (\S \ref{dim}).  In \S \ref{wavefunction}, we derive a propagation amplitude and from it a normalized wavefunction for the spatial arclength of our (1+1)-dimensional universe under the scenario of propagation from zero spatial extent (a ``Big Bang'').  In \S \ref{heatkernel} we derive the expected heat kernel by integrating the family of heat kernels on a circle of given circumference against the squared magnitude of our wavefunction, and in \S \ref{MSD} we compute the mean squared displacement of a diffusing particle according to our expected heat kernel, leading to the result that diffusion on quantum spacetime is anomalous.  Conclusions and comparisons to related results on quantum geometry in other approaches are drawn in \S \ref{discussion}.  Additionally, in \S \ref{discussion} we lay the groundwork for implementing our analysis of diffusion according to an expected heat kernel for a quantum cosmology in $(3+1)$ dimensions. 

\subsection{Quantum gravity in 1+1 dimensions}
\label{2dqg}

It is well known that in two spacetime dimensions, the field equations for general relativity vanish identically.  In light of this breakdown, an attractively geometric alternative to the curvature-minimizing Einstein-Hilbert action is Polyakov's area-minimizing action \citep{polyakov_quantum_1981} for 2d surfaces embedded in Euclidean space.  Integrating out the ``matter'' degrees of freedom represented by the embedding dimensions and working in the conformal gauge $g=e^\varphi \hat g$ for 2d metrics leads to the Liouville field theory action on the conformal degree of freedom $\varphi$.  In \citet{grumiller_liouville_2009} it is shown that Liouville gravity can be seen as a limit of Einstein gravity in $2+\varepsilon$ dimensions as $\varepsilon \to 0$.  Recent rigorous constructions \citep{david_liouville_2016,kupiainen_integrability_2020,guillarmou_conformal_2020} of Liouville quantum field theory further reaffirm the relevance of this model.

Herein we consider the reduced action
\begin{equation}
S = \int_0^T L \, dt = \int_0^T \left[ \frac{1}{4 \ell(x^0)} \left( \dot{\ell} (x^0) \right)^2 - \Lambda \ell(x^0)  \right] \, dx^0 \, ,
\label{Lagrangian}
\end{equation}
where $x^0$ is a timelike coordinate, the dynamical variable $\ell$ represents arclength around the spatial universe, and $\Lambda$ is a cosmological constant.  This action corresponds to the Hamiltonian 
\begin{align}\label{Hamiltonian} 
H=\ell \Pi_{\ell}^2 +\Lambda \ell \, ,
\end{align}
where $\Pi_{\ell} = \frac{\partial L}{\partial \dot \ell} = \frac{\dot \ell}{2 \ell}$.  In \citet{nakayama_2d_1994} the Hamiltonian \eqref{Hamiltonian} is derived from the Polyakov action by fixing a proper-time gauge, which allows application of the constraint $T_{01}=0$ to integrate spatial dependence out of $T_{00}$ and obtain the reduced Hamiltonian \eqref{Hamiltonian} after switching to Lorentzian signature.  Note we eliminate a term of the form $\frac{a}{\ell}$ added to the Hamiltonian by hand in \citet{nakayama_2d_1994} to account for Casimir energy, since in our approach an equivalent term will be generated by varying the factor ordering of the associated quantum Hamiltonian operator.  We also observe that the same action is derived as a minisuperspace model for the Liouville action in \citet{moore_loops_1991} by assuming the conformal degree of freedom to be independent of the spatial coordinate.  In \citet{moore_loops_1991} the aforementioned extra term arises in the Wheeler-DeWitt equation due to renormalized Liouville and matter couplings.

Using a Schr\"{o}dinger ansatz $\hat \ell \psi = \ell \psi(\ell)$, $\hat \Pi_{\ell} \psi = -i \hbar \frac{d \psi}{d \ell}$, the approach in \citet{nakayama_2d_1994} is to quantize the Hamiltonian \eqref{Hamiltonian} and construct propagation amplitudes for arclength $\ell$.  These constructions are extended to allow for factor ordering ambiguities and to admit a corresponding path integral definition in \citet{patel_propagators_2017, haga_factor_2017}.  Incorporating a 1-parameter family of factor orderings, we have the quantized Hamiltonian operator
\begin{align}\label{QHamiltonian} \
\begin{split}
    \hat{H} &= l^{j_1} \hat \Pi_\ell  l^{j_2} \hat \Pi_\ell l^{j_3} +\Lambda \ell , \quad j_1 + j_2 + j_3 = 1 \\
    &= -\hbar^2l^{j_1}\frac{d}{d\ell}l^{1-\left({j_3}+{j_1}\right)}\frac{d}{d \ell}l^{j_3}+\Lambda \ell \, ,
\end{split}
\end{align}
which is symmetric with respect to the measure $\ell^{j_3-j_1} d \ell$.  For convenience, we use the notation $J_\pm \equiv j_3 \pm j_1$, allowing the operator $\hat H$ to be expanded as
\begin{equation*}
\hat H = -\hbar^2 \left( \ell \frac{d^2}{d \ell^2}  + \left( 1+ J_- \right) \frac{d}{d \ell} - \frac{(J_+^2-J_-^2)}{4} \ell^{-1} \right)  + \Lambda \ell \, .
\end{equation*}

To support the computation of expectation values of observables and to construct a propagator for transition between quantum states, a quantum Hamiltonian operator must be equipped with a domain of wavefunctions on which it is self-adjoint.  Often there are infinitely many such domains to choose from.  One method of constructing the self-adjoint domains for a symmetric differential operator is by imposing asymptotic boundary conditions on functions in the natural domain of definition of its adjoint.  An integration by parts then verifies that the differential operator is self-adjoint, with the boundary terms vanishing by virtue of the imposed asymptotic behavior of domain functions (for details, see e.g. \citet{gitman_self-adjoint_2012}, Chapter 4).  

In \citet{haga_factor_2017}, it is shown that the quantized Hamiltonian \eqref{QHamiltonian} is essentially self-adjoint for $|J_+| \ge 1$, whereas for $|J_+| <1$, self-adjoint domains for \eqref{QHamiltonian} are given by the family of restrictions on asymptotic boundary behavior of wavefunctions
\begin{equation}
\psi \sim C \left(\sin(\theta) \varphi^{(1)} + \cos(\theta) \varphi^{(2)} \right) + \mathscr{O}\left( \ell^{\frac{1}{2} - \frac{J_-}{2}} \right) \, , \quad \ell \to 0 \, ,
\label{saextensions}
\end{equation}
with $\theta$ indexing the choice of self-adjoint domain, and $\varphi^{(1,2)}$ being reference modes given by (accounting for a change of variables and unitary transformation made in \citet{haga_factor_2017})
\begin{align*}
\varphi^{(1)} &= \ell^{\frac{-J_-}{2}} K_{\frac{|J_+|}{2}} \left( \frac{\sqrt{\Lambda}}{\hbar} \ell \right) \\
\varphi^{(2)} &= \ell^{\frac{-J_-}{2}} I_{\frac{|J_+|}{2}} \left( \frac{\sqrt{\Lambda}}{\hbar} \ell \right) \, ,
\end{align*}
where $I_\nu$, $K_\nu$ are modified Bessel functions.  

The factor $\ell^\frac{-J_-}{2}$ in each reference mode is present only because the Hamiltonian operator \eqref{QHamiltonian} is symmetric with respect to the weighted measure $\ell^{J_-} d \ell$ on $\R^+$.  A unitary transformation
\begin{equation}
\begin{split}
&\tilde H = U \hat H U^{-1} \\ 
&D(\tilde H) = U D(\hat H) \, ,
\end{split}
\label{unitrans}
\end{equation}
where $U \psi = \ell^\frac{J_-}{2} \psi$, yields a transformed Hamiltonian symmetric with respect to the Lebesgue measure:
\begin{align}
\begin{split}
    \tilde H \equiv \ell^{\frac{J_-}{2}} \hat H \ell^{-\frac{J_-}{2}} &= -\hbar^2 \ell^{\frac{J_+}{2}} \frac{d}{d \ell} \ell^{j_2} \frac{d}{d \ell} \ell^{\frac{J_+}{2}} + \Lambda \ell \\
    &= -\hbar^2 \ell  \frac{d^2}{d \ell^2} -\hbar^2 \frac{d}{d \ell} + \frac{\hbar^2 J_+^2}{4} \ell^{-1} + \Lambda \ell \\
    &= -\hbar^2 \frac{d}{d \ell} \ell \frac{d}{d \ell} + \frac{\hbar^2 J_+^2}{4} \ell^{-1} + \Lambda \ell \, ,
    \label{QHamsym}
\end{split}
\end{align}
where in the last line the term $\frac{\hbar^2 J_+^2}{4}\ell^{-1}$ is clearly equivalent to that added to the Hamiltonian at the classical level in \citet{nakayama_2d_1994}, as well as to the $\nu^2$ term coming from renormalized Liouville and matter couplings in the minisuperspace Wheeler-DeWitt equation given in \citet{moore_loops_1991} (see (3.13)):
\begin{equation*}
\left[ - \left( \ell \frac{\partial}{\partial \ell} \right)^2 + 4 \mu \ell^2 + \nu^2 \right] \psi = 0 \, .
\end{equation*}
Alternatively, we can proceed as in \citet{haga_factor_2017} using a similar unitary transformation $V \psi = \ell^{\frac{1}{4} + \frac{J_-}{2}} \psi$ to an operator $\tilde H '$ symmetric with respect to the measure $\ell^{-\tfrac{1}{2}} \, d \ell$.  This alternate transformation yields 
\begin{equation}
\tilde H' = - \hbar^2 \ell^{\frac{1}{2}} \frac{d}{d \ell} \ell^{\frac{1}{2}} \frac{d}{d \ell} + \frac{\hbar^2}{4} \left( J_+^2 - \frac{1}{4} \right) \ell^{-1} + \Lambda \ell \, .
\label{QHamLB}
\end{equation}
Note that the first term in $\tilde H'$ is given by the Laplace-Beltrami operator $\Delta_\ell = \ell^{\frac{1}{2}} \frac{d}{d \ell} \ell^{\frac{1}{2}} \frac{d}{d \ell}$ for the metric $g = (\ell^{-1})$ on $(0,\infty)$.  Thus the term $\frac{\hbar^2}{4} \left( J_+^2 - \frac{1}{4} \right) \ell^{-1}$ in \eqref{QHamLB} can be viewed as a quantum potential expressing the deviation of the chosen factor ordering from the Laplace-Beltrami ordering $j_1=\frac{1}{2}$, $j_3=0$.  Similarly in \eqref{QHamsym}, the quantum potential $\frac{\hbar^2 J_+^2}{4} \ell^{-1}$ represents the deviation in factor ordering from the symmetric choice $j_1=0=j_3$.  Because the operators \eqref{QHamsym} and \eqref{QHamLB} are unitarily equivalent, the choice between them is dictated by convenience for the construction at hand.

To analyze the asymptotic behavior of wavefunctions as $\ell \to 0$, it is convenient to work with the transformed Hamiltonian \eqref{QHamsym} symmetric with respect to the Lebesgue measure.  The self-adjoint extension of $\tilde H$ corresponding to $\theta = 0$ can be viewed as that in which at the singularity all wavefunctions are zero (for $|J_+| \ne 0$) or finite (for $|J_+|=0$), since $I_\nu(z) \propto z^\nu$ as $z \to 0$. 
 All other extensions contain wavefunctions approaching infinity as $\ell \to 0$, since $K_\nu(z) \propto z^{-\nu}$, $\nu > 0$, and $K_0 \propto - \log(z)$ as $z \to 0$.  In accord with these asymptotics, analysis in \citet{haga_factor_2017} identifies the $\theta=0$ domain of wavefunctions as corresponding in a path-integral approach with a choice of path-integral measure supported only on histories avoiding the singularity, while the $\theta = \frac{\pi}{2}$ domain of wavefunctions corresponds with a path-integral measure supported on histories potentially reflecting off the singularity.

For $|J_+| \ge 1$, the reference mode $\tilde \varphi^{(1)} = K_{\frac{|J_+|}{2}}(\frac{\sqrt{\Lambda} \ell}{\hbar})$ ceases to be square-integrable, demonstrating that the single self-adjoint extension of $\tilde H$ consists only of wavefunctions vanishing at the singularity.

Our wavefunction for propagation from nothing, constructed in \S \ref{wavefunction}, will belong to the self-adjoint extension corresponding to $\theta = \frac{\pi}{2}$, an unsurprising result since such a wavefunction cannot be singularity-avoiding.

\subsection{Diffusion on a circle}
\label{diffcirc}

To obtain a quantum (expected) heat kernel describing diffusion on quantized space in our model, we begin with the 1-parameter family of heat kernels for diffusion on a circle of given circumference $\ell \in (0,\infty)$.  
Using the Laplace-Beltrami operator on a circle of radius $r$ for the metric inherited from embedding in the Euclidean plane, the heat equation on a circle of radius $r$ can be written as 
\begin{equation}
\frac{1}{r^2} \partial_\omega^2 u = \partial_t u \, ,
\label{heatcirc}
\end{equation}
with periodic boundary conditions 
\begin{equation*}
u(\omega + 2 \pi) = u (\omega) \, .
\end{equation*}
Observing that in \eqref{heatcirc} the coefficient $\frac{1}{r^2}$ plays the role of a diffusion coefficient $D$ in a heat equation $D \partial_x^2 u = \partial_t u$ on Euclidean space, the heat kernel for diffusion on a circle of radius $r$ can be computed by using the heat semigroup on $\R$ and imposing periodic boundary conditions to obtain
\begin{equation*}
P_r(t, \omega) = \frac{r}{\sqrt{4 \pi t}} \sum_{k \in \Z} \exp\left(\frac{-r^2(\omega + 2\pi k)^2}{4t}\right) \, ,
\end{equation*}
or in terms of the circumference $\ell$ of the circle, 
\begin{equation}
    P_\ell(t,\omega) = \frac{\ell}{4\pi^{3/2}\sqrt{t}} \sum_{k\in\Z} \exp\left(\frac{-\ell^2(\omega+2\pi k)^2}{16\pi^2 t}\right).
    \label{heat}
\end{equation}
 
\subsection{Anomalous diffusion, walk dimension, and spectral dimension}
\label{dim}

In this section we review salient features of (anomalous) diffusion and related measures of dimension.  For a full discussion, see e.g. \citet{renner_diffusion_2005}.

A defining feature of diffusion on Euclidean space is the linear dependence of the diffusing particle's mean squared displacement on elapsed time:
\begin{equation}
\langle \Delta x^2 \rangle \propto t \, ,
\label{Fick}
\end{equation}
valid independent of spatial dimension.  Decomposing the displacement $x$ into two successive displacements $x_1+x_2 = x$ readily demonstrates that the linear relation \eqref{Fick} corresponds to independence of the individual displacements $x_1$, $x_2$, and that conversely, the dependence
\begin{equation}
\langle \Delta x^2 \rangle \propto t^\alpha , \alpha \ne 1
\label{Fickless}
\end{equation}
in anomalous diffusion is tied to a nonzero correlation between the displacements $x_1$ and $x_2$.  Anomalous diffusion with $\alpha<1$, known as \emph{subdiffusion}, has the property that on average a diffusing particle undergoes a lesser displacement over the full diffusion time $t$ than would be expected based on linearly extrapolating from its displacement over a shorter elapsed diffusion time.  Subdiffusion occurs typically in media presenting obstacles to particle motion such as a porous or fractalline structure. 

By viewing diffusion time as the time required for the diffusing particle to explore a given volume proportional to $t$ and emerge at a radius $r(t) = \langle \Delta x^2 \rangle ^{\frac{1}{2}}$, we can regard \eqref{Fickless} as giving the relation $t \propto (r(t))^{\frac{2}{\alpha}}$ between the volume of the random walk and its radial extent $r$, defining the so-called \emph{walk dimension} 
\begin{equation}
d_w=\frac{2}{\alpha}
\label{walkdim}
\end{equation}
of the path traversed by the diffusing particle.  Thus in the case of subdiffusion, the path of a diffusing particle may be seen as more intricately reticulated and hence having a higher walk dimension than the standard value of 2 on Euclidean space.

While the walk dimension is that of the diffusing particle's trajectory considered as a geometric object in its own right, properties of diffusion can also be used to investigate the dimensionality of underlying space.  The heat kernel evaluated at $x=0$, denoted $P(t,0)$, gives the return probability of a diffusing particle to its origin after diffusion time $t$, and is thus inversely related to the volume of underlying space.  Because standard diffusion in one spatial dimension is described by the Gaussian heat kernel $p(t,x) = \frac{1}{\sqrt{4\pi t}} \exp \left(-\frac{x^2}{4t} \right)$, the relative volume scaling defines the \emph{spectral dimension} of underlying space:
\begin{equation}
d_s = -2 \lim_{t \to 0} \frac{d \log P(t,0)}{d \log t} = -2 \beta \, ,
\label{specdim}
\end{equation}
assuming the scaling $P(t,0) \propto t^\beta$.

Note that by comparing the short-time scalings for radius $r(t) \propto t^{\frac{\alpha}{2}}$ and volume $V(t) \propto t^{-\beta}$ determined from diffusion, we can effectively define a dimensional exponent $\bar{d} = -\frac{2\beta}{\alpha}$ via $V \propto r^{-\frac{2\beta}{\alpha}}$, $r \to 0$, which probes the same short-distance scaling as the Hausdorff dimension $d_h = \lim_{r \to 0} \frac{\log V(r)}{\log r}$.  While we can regard $\bar{d}$ as an effective dimension of space mimicking the Hausdorff dimension, we must bear in mind that as computed here it has no direct access to the volume of a geodesic ball through length measures, but merely reconstructs this information from diffusion.  Thus it is not immediate that $\bar{d}$ must agree with the geometric definition of Hausdorff dimension.

\section{Amplitude for propagation from a Big Bang}
\label{wavefunction}

Our next objective is to construct a wavefunction for the state of a universe having propagated from zero arclength (a Big Bang) at some point in the past.  To this end, we construct a propagation amplitude $K(\ell_1, \ell_2, \tau)$ as the integral kernel for the imaginary-time evolution operator $e^{-\frac{\tau \hat{H}}{\hbar}}$ corresponding to the Hamiltonian \eqref{QHamiltonian}.  Using a spectral decomposition, the propagator can be expressed in terms of a basis $\psi_n(\ell)$ of eigenfunctions and associated eigenvalues $E_n$ of $\hat H$:
\begin{align}
K(\ell_1, \ell_2, \tau) &= \sum_{n=0}^{\infty} \psi_n(\ell_2)e^{\frac{-E_n \tau}{\hbar}}\psi_n(\ell_1) \, .
\label{spectral}
\end{align}
This approach was used in \citet{nakayama_2d_1994}; we generalize the computation therein to the case of varied factor ordering.

The eigenfunctions and eigenvalues needed for \eqref{spectral} must satisfy the differential equation 
\begin{equation}
(\hat H - E) \psi = -\hbar^2 \ell \frac{d^2\psi}{d\ell ^2}-\hbar^2\left(1+ J_- \right)\frac{d\psi}{d\ell}+\\
 \left(\hbar^2\left(\frac{J_+^2-J_-^2}{4}\ell^{-1}\right)+\Lambda \ell-E\right)\psi=0 \, ,
 \label{eigen}
\end{equation}
which is related to a confluent hypergeometric equation $zy''+(c-z)y'-ay=0$ by means of the transformation
\begin{equation*}
\psi (\ell) = e^{rz} z^s y(z) \, , \quad z = \frac{2\sqrt{\Lambda}}{\hbar} \ell \,,
\end{equation*}
with
\begin{equation*}
r = - \frac{1}{2} \, , \quad s = -\frac{-J_-}{2} \pm \frac{|J_+|}{2} \, ,
\end{equation*}
resulting in 
\begin{equation*}
a = \frac{1 \pm |J_+|}{2} - \frac{E}{2 \hbar \sqrt{\Lambda}} \, , \quad c = 1 \pm |J_+| \, .
\end{equation*}
By restricting to solutions of \eqref{eigen} in $L^2(\R^+, \ell^{J_-} d \ell )$, we obtain two orthonormal bases of eigenfunctions, distinguished by the choice $\pm$, when $|J_+|<1$:
\begin{align}\label{eigenfunctions}
\psi^{\pm}_n(\ell)=\sqrt{\frac{n! \left(\frac{2\sqrt{\Lambda}}{\hbar}\right)^{1\pm |J_{+}|}}{\Gamma\left(1\pm |J_{+}|+n\right)}}e^{-\frac{\sqrt{\Lambda} \ell}{\hbar}}\ell ^{\frac{-J_- \pm |J_+|}{2}} L_n^{(\pm |J_+|)}\left( \frac{2\sqrt{\Lambda}}{\hbar} \ell \right) \, ,
\end{align}
where the $L_n^{(\pm |J_+|)}$ are generalized Laguerre polynomials, and with corresponding eigenvalues
\begin{align}\label{eigenvalues}
E^{\pm}_n=2\hbar \sqrt{\Lambda}\left(n+\frac{1\pm |J_+|}{2}\right) \, .
\end{align}

To verify that for each choice of $\pm$ the set $\{ \psi_n^\pm \}$ forms a complete orthonormal basis of $L^2 \left( \R^+ , \ell^{J_-} d \ell \right)$, observe that under the unitary transformation \eqref{unitrans} to $L^2(\R^+ , d\ell)$, followed by a rescaling $x = \frac{2 \Lambda}{\hbar} \ell$, the eigenfunctions $\psi^\pm_n$ map to the orthonormal functions $\left( \Gamma(\alpha+1) {{n+\alpha}\choose n} \right)^{-1/2} e^{-x/2}x^{\alpha/2} L^{(\alpha)}_n(x) $ studied in \citet{szego_orthogonal_1975} (see \S5.1), with $\alpha = \pm |J_+|$.  Orthonormality results from the fact that the defining differential equation (\eqref{eigen} or its transformed version) can be written in the form of a Sturm-Liouville eigenvalue problem.  Together with appropriate asymtotic behavior near the boundary at $0$, this yields orthonormality.  For $|J_+| < 1$, the asymptotic behavior of both the $+$ and $-$ eigenbases as $\ell \to 0$ ensures the vanishing of boundary terms.  When $|J_+| \ge 1$, only the $+$ eigenbasis is square integrable.  Completeness of the system as an $L^2$ basis for $\alpha>-1$ is shown in \cite{szego_orthogonal_1975}, \S5.7.

Having obtained the eigenbases and associated eigenvalues of $\hat H$, we construct an integral kernel of the imaginary time evolution operator $e^{-\frac{\tau \hat{H}}{\hbar}}$ via spectral decomposition (one for each choice $\pm$ of the basis eigenfunctions).  Substituting (\ref{eigenfunctions}) and (\ref{eigenvalues}) into (\ref{spectral}), simplifying, and employing the Hille-Hardy formula (see e.g. \citet{erdelyi_higher_1953} \S 10.12 (20)), we get
\begin{align}
\label{propagators}
    \begin{split}
    K^{\pm}( \ell_1, \ell_2; \tau) &= \frac{\sqrt{\Lambda}}{\hbar}(\ell_1 \ell_2)^{-\frac{J_-}{2}}\csch(\sqrt{\Lambda} \tau)\exp{\left(-\frac{\sqrt{\Lambda}}{\hbar}(\ell_1+\ell_2)\coth(\sqrt{\Lambda} \tau)\right)} \,\times \\ 
    & \qquad I_{\pm |J_+|}\left(\frac{2\sqrt{\Lambda}}{\hbar} \sqrt{ \ell_1 \ell_2}\csch ( \sqrt{\Lambda} \tau )\right).
    \end{split}
\end{align}
Note, since we have an eigenbasis $\{\psi_n^+\}$ for all values of $|J_+|$ and two eigenbases $\{\psi_n^+\}$ and $\{\psi_n^-\}$ for $|J_+|<1$, spectral decomposition gives a propagator $K^+$ for all $|J_+|$ and a second propagator $K^-$ whenever $|J_+|<1$.  This situation arises because, as discussed in \S \ref{2dqg}, the Hamiltonian $\hat H$ is essentially self-adjoint for $|J_+| \ge 1$, so the propagator is uniquely defined, as promised by the functional calculus on operators resulting from the spectral theorem (see e.g. \citet{reed_functional_1980}).  When $|J_+|<1$, $\hat H$ has infinitely many self-adjoint extensions.  Similarly, for $|J_+|<1$ an infinite family of propagators is obtained by taking linear combinations
\begin{align}
\label{allpropagators}
\begin{split}
    K(\ell_1, \ell_2; \tau) &= (1-\lambda) K{^+}(\ell_1, \ell_2; \tau) + \lambda K^{-}(\ell_1, \ell_2; \tau) \ , \quad \lambda \in \mathbb{R} \,,  \\
    &= \frac{\sqrt{\Lambda}}{\hbar}(\ell_1 \ell_2)^{-\frac{J_-}{2}}\csch(\sqrt{\Lambda} \tau)\exp{\left(-\frac{\sqrt{\Lambda}}{\hbar}(\ell_1+\ell_2)\coth(\sqrt{\Lambda} \tau)\right)} \,\times \\ 
    &\qquad \Bigg[ (1-\lambda) I_{+|J_+|} \left( \frac{2\sqrt{\Lambda}}{\hbar} \sqrt{ \ell_1 \ell_2}\csch(\sqrt{\Lambda} \tau) \right) + \\
    &\pushright{\lambda I_{-|J_+|} \left( \frac{2\sqrt{\Lambda}}{\hbar} \sqrt{ \ell_1 \ell_2}\csch(\sqrt{\Lambda} \tau) \right)\Bigg]} \\
    &= \frac{\sqrt{\Lambda}}{\hbar}(\ell_1 \ell_2)^{-\frac{J_-}{2}}\csch(\sqrt{\Lambda} \tau)\exp{\left(-\frac{\sqrt{\Lambda}}{\hbar}(\ell_1+\ell_2)\coth(\sqrt{\Lambda} \tau)\right)} \,\times \\
    &\qquad \Bigg[  I_{|J_+|} \left( \frac{2\sqrt{\Lambda}}{\hbar} \sqrt{ \ell_1 \ell_2}\csch(\sqrt{\Lambda} \tau) \right) + \\
    &\pushright{\frac{2 \lambda}{\pi} \sin(\pi |J_+|) \, K_{|J_+|} \left( \frac{2\sqrt{\Lambda}}{\hbar} \sqrt{ \ell_1 \ell_2}\csch(\sqrt{\Lambda} \tau) \right) \Bigg] \, ,}
\end{split}
\end{align}
where we have used the identity $I_{-\nu}(z) - I_\nu(z) = \frac{2}{\pi} \sin(\nu \pi) K_\nu(z)$.  By observing asymptotic behavior of $K^+$ and $K^-$ as $\ell \to 0$, it is apparent that $\lambda = 1$ corresponds to the self-adjoint extension of the Hamiltonian labeled by $\theta = 0$ in \eqref{saextensions}.  Although we do not obtain a propagator from \eqref{allpropagators} corresponding to the $\theta = \frac{\pi}{2}$ self-adjoint extension in \eqref{saextensions} except formally in the limit $\lambda \to \infty$, we will see that by using any of the propagators in \eqref{allpropagators} to construct an amplitude for propagation from zero arclength at some point in the past, the resulting Big Bang wavefunction will be an element of the self-adjoint extension corresponding to $\theta = \frac{\pi}{2}$.  Notice that the necessity of taking the limit as $\lambda \to \infty$ to obtain the $\theta = \frac{\pi}{2}$ propagator also gives rise to the appearance that the case $|J_+|=0$ only yields one propagator from \eqref{allpropagators}, when in fact there are infinitely many self-adjoint extensions of the Hamiltonian for $|J_+|=0$.

In order to compute a propagation amplitude between two arclengths over an indefinite elapsed time, we integrate the time-dependent transition amplitude \eqref{propagators} over all values of elapsed time ($\tau=0$ to $\tau=\infty$).  With the change of variables $\csch(\sqrt{\Lambda} \tau) = \sinh(u)$, we can use the following formula from \citet{magnus_formulas_1966} (p.98),
\begin{align*}
I_{\nu}(az)K_{\nu}(bz) &= \int_{0}^{\infty} I_{2\nu}(2\sqrt{ab} \ z\sinh{t}) \  e^{-z(a+b)\cosh(t)} \ dt, \\
&\Re(z)>0, \quad a<b, \quad \Re(\nu) >-\frac{1}{2}.
\end{align*}
Since our integral is symmetric in $\ell_1$ and $\ell_2$, we can assume $\ell_1 < \ell_2$. With $a = \ell_1, \ b = \ell_2, \ \nu = \pm \frac{|J_+|}{2}, \ z = \frac{\sqrt{\Lambda}}{\hbar},$ and $t = u$, our propagation amplitude is
\begin{align}
    \Psi^\pm(\ell_1,\ell_2) = \int_0^\infty K^\pm \left( \ell_1 , \ell_2 ; \tau \right) \, d \tau = \frac{1}{\hbar}(\ell_1 \ell_2)^{-\frac{J_-}{2}} I_{\pm \frac{|J_+|}{2}} \left( \frac{\sqrt{\Lambda}}{\hbar}\ell_1\right)  K_{ \pm \frac{|J_+|}{2}} \left( \frac{\sqrt{\Lambda}}{\hbar}\ell_2\right) \, ,
    \label{integrated}
\end{align}
where $\Psi^+$ is valid for factor orderings with any value of $|J_+|$, while $\Psi^-$ is restricted to those with $|J_+|<1$.

\subsection{Normalizing the amplitude}

By regarding integration over elapsed time as a superposition process, we can view \eqref{integrated} as defining a (non-normalized) amplitude in $\ell$ for a universe which had definite spatial arclength $\ell_*$ at some point in the past:
\begin{equation*}
    \Psi^\pm_{\ell_*}(\ell)= \begin{cases}
    \Psi^\pm(\ell,\ell_*), & \ell < \ell_* \\
    \Psi^\pm(\ell_*,\ell), & \ell > \ell_*
    \end{cases}
\end{equation*}

To normalize over $\ell$ for fixed $\ell_*$, we must perform the integral
\begin{align}
\begin{split}
    \int_0^\infty \left\lvert \Psi_{\ell_*}^\pm(\ell) \right\rvert^2 \, \ell^{J_-} \, d\ell
    &= \frac{1}{\hbar^2}\ell_*^{-J_-} \Bigg[K^2_{\nu}\left(\tfrac{\sqrt{\Lambda}}{\hbar}\ell_*\right) \int_0^{\ell_*} I^2_{\nu}\left(\tfrac{\sqrt{\Lambda}}{\hbar}\ell\right) \, d\ell \, + \\
    &\pushright{I^2_{\nu}\left(\tfrac{\sqrt{\Lambda}}{\hbar}\ell_*\right) \int_{\ell_*}^\infty K^2_{\nu}\left(\tfrac{\sqrt{\Lambda}}{\hbar}\ell\right) \, d\ell \Bigg]} \\
    &= \frac{\ell_*^{-J_-}}{\hbar\sqrt\Lambda} \Bigg[ K^2_{\nu}\left(\tfrac{\sqrt{\Lambda}}{\hbar}\ell_*\right) \int_0^{\frac{\sqrt{\Lambda}\ell_*}{\hbar}} I^2_\nu(u) \, du + I^2_{\nu}\left(\tfrac{\sqrt{\Lambda}}{\hbar}\ell_*\right) \int_{\frac{\sqrt{\Lambda}\ell_*}{\hbar}}^\infty K^2_\nu(u) \, du \Bigg]
    \label{I1I2}
\end{split}
\end{align}
where in the last line we use the substitution $u = \frac{\sqrt\Lambda}{\hbar} \ell$.  As above, we define $\nu = \pm\tfrac{|J_+|}{2}$. We can apply various identities and substitutions to integrate \eqref{I1I2} (see \ref{appendix-normalization} for full technical details). Performing the integral yields 
\begin{align}
\begin{split}
    \int_0^\infty \lvert \Psi_{\ell_*}^\pm(\ell) \rvert^2 \, \ell^{J_-} \, d \ell &= \frac{\ell_*^{-J_-}}{2\hbar\sqrt\Lambda} \Bigg[ \frac{\left(\frac{\Lambda \ell_*^2}{\hbar^2}\right)^{\nu+\frac{1}{2}}K_\nu^2\left(\frac{\sqrt\Lambda}{\hbar}\ell_*\right)}{2^{2\nu}[\Gamma(\nu+1)]^2(\nu + \frac{1}{2})} \, \pFq{2}{3}{\nu + \frac{1}{2}, \nu + \frac{1}{2}}{\nu + \frac{3}{2},\nu + 1,2\nu + 1}{\frac{\Lambda \ell_*^2}{\hbar^2}} + \\
    & \quad \frac{\sqrt\pi}{2} \, I_\nu^2\left(\frac{\sqrt\Lambda}{\hbar} \ell_*\right) \left( \frac{\pi^{\frac{3}{2}}}{\cos(\nu\pi)} - \frac{\sqrt\Lambda}{\hbar}\ell_* \, \MeijerG*{3}{1}{2}{4}{\frac{1}{2},\frac{1}{2}}{\nu,0,-\nu,-\frac{1}{2}}{\frac{\Lambda}{\hbar^2}\ell_*^2}\right) \Bigg] \\
    &\equiv  \mathcal{N}^{-2},
    \label{pdf}
\end{split}
\end{align}
and our normalized amplitude with respect to $\ell$ becomes $\mathcal N \Psi_{\ell_*}^\pm(\ell)$.
 
To model the behavior of a universe having originated in a Big Bang, we allow $\ell_*$ to approach 0 in the normalized amplitude $\mathcal N \Psi_{\ell_*}^\pm(\ell)$.  Asymptotically as $\ell_* \to 0$, the expression given in \eqref{pdf} for $\mathcal{N}^{-2}$ will be dominated by the term $\frac{\pi^2}{2\cos(\nu\pi)} I_\nu^2\left(\frac{\sqrt\Lambda}{\hbar} \ell_*\right) $, which behaves proportionally to $\ell_*^{2 \nu}$ since $I_\nu (z) \propto z^\nu$.  Indeed, in the second term within the brackets, notice that the Meijer G-function approaches 0 as $\ell_* \to 0$ since it represents the integral of an $L^1$ function over $(0,\frac{\Lambda \ell_*^2}{\hbar^2})$ (see \eqref{termL1}). Thus as $\ell_* \to 0$ the second term will be dominated by $\frac{\pi^2}{2\cos(\nu\pi)} I_\nu^2\left(\frac{\sqrt\Lambda}{\hbar} \ell_*\right) $.  In the first term inside the square brackets, note that the generalized hypergeometric function $_2F_3$ is defined as a power series \eqref{pFq} with initial term equal to 1.  Since $K_\nu(z) \sim \frac{1}{2}\Gamma(z)(\frac{1}{2}z)^{-|\nu|}$ for $\nu \ne 0$, the first term within the brackets has asymptotic behavior proportional to $\ell_*$ for $\nu > 0$.  Similarly, when $\nu = 0$, the asymptotic behavior $K_0(z) \sim -\log(z)$ together with L'H\^opital's rule implies that the first term in the brackets approaches 0 as $\ell_* \to 0$.  For $\nu<0$, this term has asymptotic behavior proportional to $\ell_*^{1+4\nu}$, but nevertheless will still be dominated by the second term in the square brackets, since $\nu>-\frac{1}{2}$, so $1+4\nu>2\nu$.  

We conclude that as $\ell_* \to 0$, $\mathcal N \Psi_{\ell_*}^\pm(\ell)$ yields a normalized wavefunction for propagation from zero arclength:
\begin{align}
 \Psi_0(\ell) \equiv \lim_{\ell_* \to 0} \mathcal N \Psi_{\ell_*}^\pm(\ell) = \frac{2\Lambda^{\frac{1}{4}}\sqrt{\cos(\nu\pi)}}{\hbar^{\frac{1}{2}}\pi} \, \ell^{-\frac{J_-}{2}} K_\nu\left(\frac{\sqrt\Lambda}{\hbar} \ell\right).
\label{BBwavefunction}
\end{align}
Note that for $ \lvert \nu \rvert \ge \frac{1}{2}$, we cannot normalize a wavefunction for propagation from 0, because $K_\nu(z) \sim \frac{1}{2}\Gamma(|\nu|)(\frac{1}{2}z)^{-|\nu|}$.  

For purposes of comparing wavefunctions with differing factor orderings, we observe as in \S \ref{2dqg} that the factor $\ell^{-\frac{J_-}{2}}$ in \eqref{BBwavefunction} is present only because the Hamiltonian \eqref{QHamiltonian} is symmetric with respect to the measure $\ell^{J_-} \, d \ell$, and hence $\Psi_0 \in L^2 \left( \R^+ , \ell^{J_-} d \ell \right)$.  However \eqref{QHamiltonian} is unitarily equivalent to the operator \eqref{QHamsym}, symmetric with respect to the Lebesgue measure.  Multiplication by $\ell^{\frac{J_-}{2}}$ as in \eqref{unitrans} correspondingly transforms wavefunctions to be normalized with respect to the Lebesgue measure, so that from \eqref{BBwavefunction} we have 
\begin{equation}
\tilde \Psi_0(\ell) \equiv \frac{2\Lambda^{\frac{1}{4}}\sqrt{\cos(\nu\pi)}}{\hbar^{\frac{1}{2}}\pi} \,  K_\nu\left(\frac{\sqrt\Lambda}{\hbar} \ell\right).
\label{BBwavefunctionsym}
\end{equation}
Because $K_{-\nu}(z) = K_\nu(z)$, the wavefunction \eqref{BBwavefunctionsym} is independent of the sign of $\nu$, and accordingly we can henceforward assume that $\nu = \frac{|J_+|}{2}$.  Relatedly, we note that \eqref{BBwavefunction} is an element of the self-adjoint extension \eqref{saextensions} of $\hat H$ corresponding to $\theta = \frac{\pi}{2}$ (indeed, \eqref{BBwavefunction} is a multiple of the reference mode $\varphi^{(1)}$ (see \S \ref{2dqg})).  As discussed in \S \ref{2dqg} and in \citet{haga_factor_2017}, the $\theta = 0$ self-adjoint extension of $\hat H$ corresponds to a singularity-avoiding ansatz, so it is to be expected that a wavefunction for propagation from the singularity cannot be defined for orderings with $|J_+| \ge 1$, which only allow singularity-avoiding scenarios. 

Although we have constructed \eqref{BBwavefunction} in terms of a propagation amplitude from zero arclength, it is also a solution to the Wheeler-DeWitt equation $\hat H \Psi = 0$.  In fact it is the only smooth normalizable solution, since the scaling $z=\frac{\sqrt{\Lambda}}{\hbar} \ell$ transforms the Lebesgue-symmetrized Wheeler-DeWitt equation $\tilde H \tilde \Psi = 0$ (see \eqref{QHamsym}) to be equivalent to the modified Bessel equation $z^2 u'' + z u' - \left( z^2 + \nu^2 \right)u = 0$, whose solution space is spanned by $K_\nu(z)$ and $I_\nu(z)$ (the amplitude $\Psi_{\ell_*}^\pm(\ell)$ is smooth on $\R^+ \setminus \{ \ell_* \}$).  Asymptotically as $z \to \infty$, $K_\nu(z)$ and $I_\nu(z)$ behave proportionally to $z^{-\frac{1}{2}} e^{-z}$ and $z^{-\frac{1}{2}} e^z$, respectively, making $K_\nu(z)$ the only smooth solution with any hope of having a finite $L^2$ norm.  Normalizability is allowed only in the range $\lvert \nu \rvert < \frac{1}{2}$ by the asymptotics $K_\nu(z) \propto z^{-\nu}$, $K_0(z) \propto -\log(z)$ as $z$ tends to 0.  Thus outside the range of orderings $\lvert J_+ \rvert <1$, the Wheeler-DeWitt equation has no normalizable solution.

\section{Quantum heat kernel}
\label{heatkernel}

Next, we use the wavefunction obtained in the previous section to compute the expected heat kernel describing diffusion in our model universe. The heat kernel $P_\ell(t,\omega)$ on a circle of circumference $\ell$ is given in \eqref{heat} by
\begin{equation*}
    P_\ell(t,\omega) = \frac{\ell}{4\pi^{\frac{3}{2}}\sqrt{t}} \sum_{k\in\Z} \exp\left(\frac{-\ell^2(\omega+2\pi k)^2}{16\pi^2 t}\right).
\end{equation*}
Thus, to compute the quantum (expected) heat kernel $P(t,\omega)$ we must integrate \eqref{heat} against the squared magnitude of the wavefunction \eqref{BBwavefunctionsym}. Since all summands in our integrand are nonnegative, we can apply the monotone convergence theorem to get
\begin{align*}
    P(t,\omega) &\equiv \int_0^\infty P_\ell(t,\omega) \left| \tilde{\Psi}_0(\ell)\right|^2 d\ell \\
    &= \frac{\sqrt\Lambda \cos(\nu\pi)}{\hbar\pi^{\frac{7}{2}} \sqrt t} \sum_{k \in \Z}\int_0^\infty \ell \exp\left(\frac{-\ell^2(\omega+2\pi k)^2}{16\pi^2 t}\right) K_\nu^2\left(\frac{\sqrt\Lambda}{\hbar}\ell\right) \, d\ell.
\end{align*}
Making the substitution $u = \frac{\sqrt\Lambda}{\hbar}\ell$ yields
\begin{equation}
    P(t,\omega) = \frac{\hbar\cos(\nu\pi)}{\pi^{\frac{7}{2}}\sqrt{\Lambda t}} \sum_{k \in \Z} \int_0^\infty u\exp\left(\frac{-u^2\hbar^2(\omega+2\pi k)^2}{16\pi^2\Lambda t}\right) K_\nu^2(u) \, du.
    \label{qheatint}
\end{equation}
Now we are able to use (10) from \citet{ragab_product_1955} (see page 126) to evaluate the integral by taking $l = 2$, $m = \nu = n$, and $z = \frac{16\pi^2\Lambda t}{\hbar^2(\omega+2\pi k)^2}$ (assuming that $k$ and $\omega$ are not both zero, a case to be considered separately).  Hence, we have
\begin{equation*}
    P(t,\omega) = \frac{\hbar\cos(\nu\pi)}{4\pi^3\sqrt{\Lambda t}} \sum_{k \in \Z} \MeijerG*{4}{1}{3}{4}{1,1,\frac{3}{2}}{1+\nu,1,1,1-\nu}{\frac{16\pi^2\Lambda t}{\hbar^2(\omega+2\pi k)^2}}.
\end{equation*}
Finally, by \eqref{GReduction}, we can reduce the Meijer G-functions in the sum to obtain the quantum heat kernel
\begin{equation}
    P(t,\omega) = \frac{\hbar\cos(\nu\pi)}{4\pi^3\sqrt{\Lambda t}} \sum_{k\in \Z} \MeijerG*{3}{1}{2}{3}{1,\frac{3}{2}}{1+\nu,1,1-\nu}{\frac{16\pi^2\Lambda t}{\hbar^2(\omega+2\pi k)^2}}\,, \quad \omega \ne 0.
    \label{qheat}
\end{equation}

In the case where $\omega=0$, the $k=0$ term of \eqref{qheatint} reduces to the following, much simpler integral:
\begin{equation*}
    \frac{\hbar\cos(\nu\pi)}{\pi^{\frac{7}{2}}\sqrt{\Lambda t}} \int_0^\infty u K_\nu^2(u) \, du \, ,
\end{equation*}
whose integrand can be rewritten using \eqref{KtoG} with $\omega=1$ and $\mu = \nu$. Thus, the integral becomes
\begin{equation*}
    \frac{\hbar\cos(\nu\pi)}{2\pi^3 \sqrt{\Lambda t}} \int_0^\infty \MeijerG*{3}{0}{1}{3}{1}{\frac{1}{2}+\nu,\frac{1}{2},\frac{1}{2}-\nu}{u^2} \, du \,
\end{equation*}
which can be evaluated using the substitution $z = u^2$, \eqref{MeijerDefInt}, and \eqref{GIntCase2}. We set $s = \frac{1}{2}$, $m=3$, $n=0$, $p=1$, and $q=3$, so $\delta = 1$. The verification of the necessary conditions as outlined in \eqref{GIntCase2} is straightforward. Hence, applying \eqref{MeijerDefInt} yields
\begin{equation*}
    \int_0^\infty \MeijerG*{3}{0}{1}{3}{1}{\frac{1}{2}+\nu,\frac{1}{2},\frac{1}{2}-\nu}{u^2} \, du
    = \frac{2\Gamma(1+\nu)\Gamma(1-\nu)}{\sqrt\pi}.
\end{equation*}
For the $k=0$ term of $P(t,0)$, we thus have
\begin{equation*}
    \frac{\hbar\cos(\nu\pi) \Gamma(1+\nu)\Gamma(1-\nu)}{2\pi^{\frac{7}{2}} \sqrt{\Lambda t}}.
\end{equation*}
Note that the case where $\nu = 0$ simplifies to 
\begin{equation*}
\frac{\hbar}{2 \pi^{\frac{7}{2}} \sqrt{\Lambda t}},
\end{equation*}
whereas when $\nu \neq 0$ we can use the definition of the gamma function and Euler's reflection formula to obtain
\begin{equation*}
    \frac{\hbar\nu\cot(\nu\pi)}{2\pi^{\frac{5}{2}}\sqrt{\Lambda t}} ,
\end{equation*}
which can readily be verified to match the value for $\nu=0$ in the limit. 

Hence for $\omega = 0$, we have the complete expression for the quantum heat kernel $P(t,0)$ as
\begin{equation}
    P(t,0) = 
    \frac{\hbar\nu\cot(\nu\pi)}{2\pi^{\frac{5}{2}}\sqrt{\Lambda t}} + \frac{\hbar\cos(\nu\pi)}{4\pi^3\sqrt{\Lambda t}} \sum_{k \in \Z \setminus \{0\}} \MeijerG*{3}{1}{2}{3}{1,\frac{3}{2}}{1+\nu,1,1-\nu}{\frac{4 \Lambda}{\hbar^2 k^2} t } \, ,
    \label{Pt}
\end{equation}
where for $\nu=0$ we give the first term its limiting value of $\frac{\hbar}{2 \pi^{\frac{7}{2}} \sqrt{\Lambda t}}$.

\subsection{Return Probability from Asymptotics of the Quantum Heat Kernel}
\label{return}

The quantum heat kernel \eqref{Pt} for $\omega=0$ obtained above can be interpreted as the probability that, after diffusion time $t$, the particle returns to its original position $\omega=0$.

In examples of diffusion on a Riemannian manifold, this return probability can be expressed as a power series in $t$ with a prefactor of order $t^{-\frac{d}{2}}$, where $d$ is the spectral dimension of space, coinciding for Riemannian manifolds with the topological dimension.  In the case of diffusion on a Riemannian manifold with boundary, terms of half-integer order appear in the power series expansion \citep{vassilevich_heat_2003}.

In this subsection we investigate the behavior of return probability from the quantum heat kernel \eqref{Pt}, finding that the spectral dimension retains its classical value of 1, but that the subsequent expansion of the return probability is modified from the power series form expected on a Riemannian manifold.  These modifications deviate more strongly from power series behavior as $\nu$ increases. 
 The estimation of the return probability proceeds somewhat differently in the $\nu \ne 0$ and the $\nu = 0$ cases, which are treated below.

\subsubsection{Return Probability for $\nu \ne 0$}

To evaluate the asymptotic behavior of \eqref{Pt} when $\nu$ is nonzero, we use the expansion of the Meijer G-function in terms of generalized hypergeometric functions, along with identities for the gamma function (the defining identity, Euler reflection formula, and Legendre duplication formula), obtaining
\begin{align}
\begin{split}
    P(t,0) = &\frac{\hbar \cot(\nu \pi)}{2 \pi^{\frac{5}{2}} \sqrt{\Lambda t}} \left[ \nu + \sum_{k \in \Z \setminus \{0\}} \left( \frac{\Gamma ( \nu )}{4^{1-\nu}} (\rho t)^{1-\nu} \pFq{1}{1}{\frac{1}{2}-\nu}{1-2\nu}{\rho t} - \right. \right. \\
    & \pushright{\frac{1}{2\nu} (\rho t) \pFq{2}{2}{1,\frac{1}{2}}{1-\nu,1+\nu}{\rho t} - \frac{\Gamma ( -\nu)}{4^{1+\nu}} (\rho t)^{1+\nu} \pFq{1}{1}{\frac{1}{2}+\nu}{1+2\nu}{\rho t} \Bigg) \Bigg] \, ,}
    \label{Ptbeta}    
\end{split}
\end{align}
where for convenience we have set $\rho = \frac{4 \Lambda}{\hbar^2 k^2}$ .  The case $\nu=0$ will be considered separately later in this section.

From the form \eqref{Ptbeta}, several features become apparent.  Asymptotically $P(t,0) \propto t^\beta$ with $\beta = -\frac{1}{2}$ for small $t$, implying that the spectral dimension \eqref{specdim} of quantized space in our model is $d_s=1$, unchanged from its classical value.  However, the effects of quantization on spatial geometry as manifested in diffusion are evident in two ways.  First, comparing our result
\begin{equation*}
    P(t,0) \sim \frac{\hbar \nu \cot(\nu \pi)}{2 \pi^{\frac{5}{2}} \sqrt{\Lambda t}} \, , \quad t \to 0
\end{equation*}
for $\nu\ne 0$ with the Euclidean case 
\begin{equation*}
    P(t,0) \sim \frac{1}{\sqrt{4 \pi D t}} \, , \quad t \to 0 \, , 
\end{equation*}
where $D$ is the diffusion coefficient, we see that the effect of $\nu$ increasing toward $\frac{1}{2}$ (recall that $\nu$ can assume any value in $[0,\frac{1}{2})$) is analogous to that of the diffusion coefficient in the Euclidean case growing toward infinity.  The larger the value of $\nu$, the lesser the chance of the diffusing particle to return to its original position, because diffusion carries it down a path of (almost) no return.  Subsequent sections will shed further light on this result by revealing that the dimension of the random walk pursued by the diffusing particle increases with $\nu$, becoming infinite as $\nu$ tends toward $\frac{1}{2}$.

As a second effect of quantization for $\nu \ne 0$ evidenced by the dependence of \eqref{Ptbeta} on $\nu$, observe that the generalized hypergeometric functions are defined as power series, and because of the prefactors on the $_1F_1$ terms, the summation for $k \ne 0$ includes terms with noninteger power, suggesting that diffusion proceeds differently from on a Riemannian manifold.

\subsubsection{Return Probability for $\nu = 0$}
Analyzing the expansion of the return probability in the case $\nu = 0$ is slightly more involved, as the expression of the Meijer G-functions in terms of generalized hypergeometric functions no longer applies:  the poles in the contour integral that defines the Meijer G-function are no longer simple, complicating the application of the residue theorem. Hence, to examine the higher-order terms ($k \ne 0$) in the expansion \eqref{Pt} we must directly use the definition \eqref{MeijerGDef} for a Meijer G-function, yielding
\begin{equation}
    \MeijerG*{3}{1}{2}{3}{1,\frac{3}{2}}{1,1,1}{z} = \frac{1}{2\pi i} \int_L \frac{(\Gamma(1-s))^3 \Gamma(s)}{\Gamma(\frac{3}{2}-s)} z^s \, ds
    \label{nu=0_G_integral}
\end{equation}
where $L$ is a loop beginning and ending at $+\infty$, encircling all poles of $\Gamma(b_j - s)$, $j=1,2,3$, exactly once in the negative direction, but not encircling any pole of $\Gamma(1-a_k+s)$, $k=1$ (for a discussion of the choice of contour in the definition of $G^{m,n}_{p,q}$ depending on the values of $p$ and $q$, see e.g. \citet{luke_special_1969}, \S5.2). Since the poles of the gamma function occur at non-positive integer values, we have the poles of $\Gamma(b_j - s)$ ($b_j = 1$, $j = 1,2,3$) at $s = 1,2,3,\dots$. Note that the poles of $\Gamma(z)$ are simple, as can be seen from repeatedly applying the recurrence formula $z\Gamma(z) = \Gamma(z+1)$, so the poles of $(\Gamma(1-s))^3$ are of order 3. 

To compute the residue at $s=1$, we use the recurrence formula of the gamma function, specifically $(1-s)\Gamma(1-s) = \Gamma(2-s)$, to rewrite the integrand, and then the residue theorem to obtain
\begin{equation*}
    \frac{1}{2\pi i} \int_L \frac{(\Gamma(1-s))^3 \Gamma(s)}{\Gamma(\frac{3}{2}-s)} z^s \, ds = - \sum_{a \in \N} \mathop{\mathrm{Res}}_{s=a} \left[\frac{(\Gamma(2-s))^3 \Gamma(s)}{(1-s)^3\Gamma(\frac{3}{2} - s)} z^s\right].
\end{equation*}
Denote by $g_1(s)$ the function $\frac{(\Gamma(2-s))^3 \Gamma(s)}{\Gamma(\frac{3}{2}-s)}$ and observe that $g_1(s)$ is analytic at $s=1$. Furthermore, notice that $z^s$ is analytic at $s=1$ for $z\neq 0$ as long as a branch is specified for $\log(z)$. Thus, we have
\begin{align*}
    -\mathop{\mathrm{Res}}_{s=1} \left[\frac{(\Gamma(2-s))^3 \Gamma(s)}{(1-s)^3\Gamma(\frac{3}{2} - s)} z^s\right] &= \mathop{\mathrm{Res}}_{s=1} \left[\frac{g_1(s)z^s}{(s-1)^3}\right] = \frac{1}{2!} \frac{d^2}{ds^2} \left[g_1(s)z^s\right]\Big|_{s=1} \\
    &= \frac{z}{2} \left(g_1''(1) + 2g_1'(1)\log(z) + g_1(1)(\log(z))^2\right).
\end{align*}

To compute the residue term at $s=2$, we further rewrite the integrand 
in \eqref{nu=0_G_integral} using $(2-s)\Gamma(2-s) = \Gamma(3-s)$, so that
\begin{align*}
    -\mathop{\mathrm{Res}}_{s=2} \left[\frac{(\Gamma(2-s))^3 \Gamma(s)}{(1-s)^3\Gamma(\frac{3}{2} - s)} z^s\right] &= -\mathop{\mathrm{Res}}_{s=2} \left[ \frac{(\Gamma(3-s))^3 \Gamma(s)}{(2-s)^3(1-s)^3\Gamma(\frac{3}{2} - s)} z^s \right] \\
    &= \mathop{\mathrm{Res}}_{s=2} \left[\frac{g_2(s)z^s}{(s-2)^3}\right] \\
    &= \frac{z^2}{2}\left(g_2''(2) + 2g_2'(2)\log(z) + g_2(2)(\log(z))^2\right) \,
\end{align*}
where by $g_2(s)$ we denote the function $\frac{(\Gamma(3-s))^3 \Gamma(s)}{(1-s)^3\Gamma(\frac{3}{2}-s)}$.

We can continue this process for the subsequent residues. Notice that what remains is not a standard power series but instead a power series modified by logarithmic terms.  It is easily verified, for example by a computation in Mathematica, that the coefficients $g_1(1)$, $g_1'(1)$, $g_1''(1)$, and $g_2(2)$, $g_2'(2)$, $g_2''(2)$ are nonzero.  Thus for both $\nu=0$ and $\nu \ne 0$, the expansion \eqref{Pt} for return probability is modified from the power series form common to diffusion on a Riemannian manifold.  Because both $z^n \log(z)$ and $z^n (\log(z))^2$ are $o(z^{n-\nu})$ as $z \to 0$, the modifications are most moderate in the $\nu=0$ case, and grow more significant as $\nu$ increases toward $\frac{1}{2}$.

To gain insight, note that in terms of the wavefunction for our Big-Bang universe, the factor ordering parameter $\nu = \frac{|J_+|}{2}$ quantifies the degree to which amplitudes are biased toward universes of small spatial arclength.  As seen at the end of \S \ref{wavefunction}, our wavefunction for propagation from nothing (transformed to be symmetric with respect to the Lebesgue measure) is given by \eqref{BBwavefunctionsym}:
\begin{equation*}
\tilde \Psi_0(\ell) = \frac{2\Lambda^{\frac{1}{4}}\sqrt{\cos(\nu\pi)}}{\hbar^{\frac{1}{2}}\pi} \,  K_\nu\left(\frac{\sqrt\Lambda}{\hbar} \ell\right) \, .
\end{equation*}
According to the asymptotics \eqref{Besselasymp} of the Bessel function $K_\nu(z)$, we have
\begin{equation}
\tilde \Psi_0(\ell) \propto \begin{cases} \ell^{-\nu} \, , & \nu \ne 0 \\
- \log \left( \frac{\sqrt{\Lambda}}{\hbar} \ell \right) \, ,  & \nu = 0  \end{cases} \, ,
\label{waveasymp}
\end{equation}
so the larger the value of $\nu$, the more sharply the amplitude $\tilde \Psi_0$ rises as $\ell \to 0$.  Hence increased digression of the return probability's expansion from power-series form is linked to stronger bias of the wavefunction toward small spatial arclengths.
% Hence for factor orderings where the Big-Bang universe wavefunction tends toward small spatial circumference, the effect of quantization on the diffusion process becomes more extreme. 
\section{Mean squared displacement}
\label{MSD}

In this section, by computing the mean squared displacement of a particle diffusing according to the quantum heat kernel, we probe more deeply into the non-manifoldlike character of diffusion on our quantized space, finding that it is indeed anomalous, with
\begin{equation}
\langle \Delta \omega^2 \rangle \propto \begin{cases} t^{\frac{1}{2} - \nu} \, , & 0 < \nu < \frac{1}{2} \\
t^{\frac{1}{2}} \left( \log(t) \right)^2 \, , & \nu = 0 \, . \end{cases}
\label{diffalpha}
\end{equation}
Thus for all $\nu \in [0,\frac{1}{2})$, diffusion over elapsed time $t$ has a smaller mean squared displacement than would be predicted by extrapolating linearly over a subinterval. 

For nonzero $\nu$, the anomalous diffusion \eqref{diffalpha} matches the typical subdiffusive behavior observed on porous materials and fractals.  The intuition in such cases is that the medium presents obstacles which act cumulatively to restrict the expected displacement of a diffusing particle.  For $\nu=0$ the asymptotic behavior of $\langle \Delta \omega^2 \rangle$ is generalized by the presence of the logarithmic factor to what \citet{oliveira_anomalous_2019} term `weak subdiffusion' (although in \citet{oliveira_anomalous_2019} it is considered in the limit of long diffusion time).  The asymptotic behavior seen in \eqref{diffalpha} accords with that of the wavefunction $\tilde \Psi_0 \left( \ell \right)$ (see \eqref{waveasymp}). As in the consideration of return probabilities (\S \ref{return}), we see that the anomalous diffusion behavior \eqref{diffalpha} becomes more pronounced as $\nu$ increases toward $\frac{1}{2}$. 

\subsection{Mean squared displacement for $\nu \ne 0$}
To compute $\langle \Delta \omega^2 \rangle$, we integrate $\omega^2$ against the quantum heat kernel:
\begin{equation*}
\langle \Delta \omega^2 \rangle = \frac{\hbar \cos(\nu \pi)}{4 \pi^3 \sqrt{\Lambda t}} \int_{-\pi}^\pi \omega^2 \sum_{k \in \Z} \MeijerG*{3}{1}{2}{3}{1 , \frac{3}{2}}{1+\nu , 1, 1-\nu}{\frac{16\Lambda \pi^2 t}{\hbar^2 (\omega+2\pi k)^2}} \, d\omega \, .
\end{equation*}
All terms in the summation defining the quantum heat kernel are nonnegative since they result from integration of nonnegative functions, so the monotone convergence theorem justifies interchanging the integral and sum.  Additionally, separating the $k=0$ term from the sum, we have
\begin{align}
\begin{split}
    \langle \Delta \omega^2 \rangle = \frac{\hbar \cos(\nu \pi)}{4 \pi^3 \sqrt{\Lambda t}} & \left[ 2 \int_0^\pi \omega^2 \MeijerG*{3}{1}{2}{3}{1 , \frac{3}{2}}{1+\nu , 1, 1-\nu}{\frac{16\Lambda \pi^2 t}{\hbar^2 }\omega^{-2}} \, d\omega \right. + \\
    & \pushright{\sum_{k \in \Z \setminus \{0\}} \int_{-\pi}^\pi \omega^2 \MeijerG*{3}{1}{2}{3}{1 , \frac{3}{2}}{1+\nu , 1, 1-\nu}{\frac{16\Lambda \pi^2 t}{\hbar^2 }(\omega+2\pi k)^{-2} } \, d\omega \Bigg] \, .}
\end{split}
\label{MSD_int}
\end{align}
For the first integral, the substitution $u=\frac{16 \Lambda \pi^2 t}{\hbar^2} \omega^{-2}$ and the integration formula \eqref{MeijerIndefInt}, together with the identity \eqref{MeijerPowerProd}, yield
\begin{align*}
    \frac{\hbar \cos(\nu \pi)}{2 \pi^3 \sqrt{\Lambda t}} \int_0^\pi \omega^2 &\MeijerG*{3}{1}{2}{3}{1 , \frac{3}{2}}{1+\nu , 1, 1-\nu}{\frac{16\Lambda \pi^2 t}{\hbar^2 }\omega^{-2}} \, d\omega \\
    &\pushright{=\frac{16 \Lambda t \cos(\nu \pi )}{\hbar^2} \left[ - \left. \MeijerG*{3}{1}{2}{3}{-\frac{1}{2} , 1}{-\frac{1}{2}+\nu , -\frac{1}{2}, -\frac{1}{2}-\nu}{u} \right|^\infty_{\frac{16\Lambda t}{\hbar^2}} \right] \, .}
\end{align*}
To evaluate the limit of the Meijer G-function at infinity, we use results detailed in \citet{luke_special_1969} for the asymptotic expansion of Meijer G-functions in terms of (divergent) series which formally define generalized hypergeometric functions (see Theorem 1, \S5.7).  In this case, the leading-order asymptotic expansion implies that in the limit as $u$ tends to infinity, $\MeijerG*{3}{1}{2}{3}{-\frac{1}{2} , 1}{-\frac{1}{2}+\nu , -\frac{1}{2}, -\frac{1}{2}-\nu}{u}$ decays like $u^{-3/2}$, so that the first integral in \eqref{MSD_int} becomes
\begin{equation}
\frac{16 \Lambda t \cos(\nu \pi )}{\hbar^2} \MeijerG*{3}{1}{2}{3}{-\frac{1}{2} , 1}{-\frac{1}{2}+\nu , -\frac{1}{2}, -\frac{1}{2}-\nu}{\frac{16\Lambda t}{\hbar^2}} \, .
\label{firstterm}
\end{equation}
To quantify the dependence of the Meijer G-function on diffusion time $t$, for $\nu \ne 0$ we express it in terms of generalized hypergeometric functions, using the formula \eqref{MeijerG_F} (the case $\nu=0$ will be treated in the following subsection).  Simplifying the result, using well-known identities for the gamma function (its functional relation $z \Gamma (z) = \Gamma (z+1)$ along with the Euler reflection formula $\Gamma (1-z) \Gamma (z) = \frac{\pi}{\sin(\pi z)}$ (valid for $z \notin \Z$) and the Legendre duplication formula $\Gamma(z) \Gamma(z+\frac{1}{2}) = 2^{1-2z} \sqrt \pi \Gamma (2z)$ (valid for $z \ne -\frac{n}{2}$, $n=0,1,2,\dots$))
we obtain
\begin{align*}
    \frac{16 \Lambda t \cos(\nu \pi )}{\hbar^2} &\MeijerG*{3}{1}{2}{3}{-\frac{1}{2} , 1}{-\frac{1}{2}+\nu , -\frac{1}{2}, -\frac{1}{2}-\nu}{Ct}  \\
    &= \frac{\sqrt \pi}{\tan(\nu \pi)} \left( Ct \right)^{\frac{1}{2}} \Bigg\{ \frac{2^{2 \nu} \Gamma(\nu)}{1+2\nu} \left( Ct \right)^{-\nu} \pFq{1}{1}{-\frac{1}{2}-\nu}{1-2\nu}{Ct} -
    \\
    &\pushright{\qquad\frac{2}{\nu} \pFq{2}{2}{1,-\frac{1}{2}}{1-\nu,1+\nu}{Ct} - \frac{2^{-2 \nu} \Gamma(-\nu)}{1-2 \nu} \left( Ct \right)^{\nu} \pFq{1}{1}{-\frac{1}{2}+\nu}{1+2\nu}{Ct} \Bigg\} \, ,}
\end{align*}
where $C=\frac{16 \Lambda}{\hbar^2}$.  Since each hypergeometric function is defined by a power series expansion, for small $t$ the dominant behavior of this term will be of the order $ t^{\frac{1}{2} - \nu}$.

Turning our attention to the $k \in \Z \setminus \{0 \}$ terms in the sum \eqref{MSD_int}, we note first that similarly to the case of the $k=0$ term, the Meijer G-function can be reformulated in terms of generalized hypergeometric functions using \eqref{MeijerG_F}:
\begin{align}
\begin{split}
    \MeijerG*{3}{1}{2}{3}{1 , \frac{3}{2}}{1+\nu , 1, 1-\nu}{\rho t} &= \frac{\Gamma(-\nu) \Gamma(-2\nu) \Gamma(1+\nu)}{\Gamma \left( \frac{1}{2}-\nu \right)} ( \rho t)^{1+\nu} \pFq{1}{1}{\frac{1}{2}+\nu}{1+2\nu}{ \rho t} + \\
    &\quad\frac{\Gamma(\nu) \Gamma(-\nu)}{\sqrt{\pi}} ( \rho t) \pFq{2}{2}{1, \frac{1}{2}}{1-\nu , 1+ \nu}{\rho t} + \\
    &\quad\frac{\Gamma(2\nu) \Gamma(\nu) \Gamma(1-\nu)}{\Gamma \left( \frac{1}{2}+\nu \right)} ( \rho t)^{1-\nu} \pFq{1}{1}{\frac{1}{2}-\nu}{1-2\nu}{ \rho t},
    \label{Gknonzero}    
\end{split}
\end{align}
where for simplicity we set $\rho=\frac{16 \Lambda \pi^2 }{\hbar^2} (\omega+2 \pi k)^{-2}$ (note that because $k \in \Z \setminus \{0\}$ and $\omega \in (-\pi, \pi]$, $\rho$ is well defined).  Each generalized hypergeometric function in \eqref{Gknonzero} is defined by a power series expansion (see \eqref{pFq}) having first term equal to 1.  Thus
\begin{align*}
    \frac{\hbar \cos(\nu \pi)}{4 \pi^3 \sqrt{\Lambda t}} \int_{-\pi}^\pi \omega^2 \MeijerG*{3}{1}{2}{3}{1 , \frac{3}{2}}{1+\nu , 1, 1-\nu}{\rho t } \, d \omega &= C_1 t^{\frac{1}{2}-\nu} \int_{-\pi}^\pi \omega^2 \rho^{1-\nu} \left( 1 + \mathscr{O} (\rho t) \right) d\omega \, + \\
    &\quad C_2 t^{\tfrac{1}{2}} \int_{-\pi}^\pi \omega^2 \rho \left( 1 + \mathscr{O} (\rho t) \right) d\omega \, + \\
    &\quad C_3 t^{\frac{1}{{2}}+\nu} \int_{-\pi}^\pi \omega^2 \rho^{1+\nu} \left( 1 + \mathscr{O} (\rho t) \right) d\omega  \, ,
\end{align*}
where the constants $C_1$, $C_2$, and $C_3$ depend on $\nu$, $\Lambda$, $\pi$, and $\hbar$.  We conclude that each term in \eqref{MSD_int} is of the order $t^{\frac{1}{2}- \nu}$ for small $t$, so that we obtain the asymptotic behavior
\begin{equation*}
\langle \Delta \omega ^2 \rangle \propto t^{\frac{1}{2} - \nu} \, , \quad t \to 0\,, \quad \nu \ne 0 \, ,
\end{equation*}
indicating that diffusion according to the expected heat kernel in the present model of (1+1)-dimensional quantum gravity is anomalous, with the exponent $\alpha = \frac{1}{2} - \nu$ dependent on the factor ordering.  Recall $\nu \in [ 0,\frac{1}{2} )$, so with the current restriction of $\nu \ne 0$, the exponent $\alpha = \frac{1}{2} - \nu$ satisfies $0<\alpha < \frac{1}{2}$.

\subsection{Mean squared displacement for $\nu = 0$}
Because the lower indices in the Meijer G-functions appearing in the quantum heat kernel are all equal for $\nu=0$, the poles in the contour integral defining each Meijer G-function are of order 3, and the application of the residue theorem must be revisited, similarly to our calculation of return probabilities for $\nu=0$ in \S \ref{return}.

Integrating the first term in \eqref{MSD_int} leads to \eqref{firstterm}, which with $\nu=0$ becomes
\begin{equation}
 C t \cdot \MeijerG*{3}{1}{2}{3}{-\frac{1}{2} , 1}{-\frac{1}{2} , -\frac{1}{2}, -\frac{1}{2}}{Ct} = Ct \cdot \frac{1}{2\pi i} \int_L \frac{ \left( \Gamma \left( - \frac{1}{2} - s \right) \right)^3 \Gamma \left( \frac{3}{2} + s \right) }{\Gamma \left( 1-s \right)} \left( Ct \right)^s \, ds  \, ,
\label{firsttermnu=0}
\end{equation}
where as before $C=\frac{16 \Lambda}{\hbar^2}$, and $L$ is a contour beginning and ending at $+\infty$ that encircles all the poles of $\Gamma \left( - \frac{1}{2} -s \right)$ (at $s = -\frac{1}{2}, \, \frac{1}{2} , \, \frac{3}{2} , \, \dots$) exactly once, and does not encircle any of the poles of $\Gamma \left( \frac{3}{2}+s \right)$ (at $s=-\frac{3}{2}, \, -\frac{5}{2} , \, \dots$).  The residue theorem implies 
\begin{align*}
\frac{1}{2\pi i} \int_L \frac{ \left( \Gamma \left( - \frac{1}{2} - s \right) \right)^3 \Gamma \left( \frac{3}{2} + s \right) }{\Gamma \left( 1-s \right)} (Ct)^s \, ds &=  - \sum_{a=0}^\infty  \left( \mathop{\mathrm{Res}}_{s=a-\frac{1}{2}} \frac{ \left( \Gamma \left( - \frac{1}{2} - s \right) \right)^3 \Gamma \left( \frac{3}{2} + s \right) }{\Gamma \left( 1-s \right)} (Ct)^s \right) \\
&= \sum_{a=0}^\infty  \left( \mathop{\mathrm{Res}}_{s=a-\frac{1}{2}} \frac{ \left( \Gamma \left(\frac{1}{2} - s \right) \right)^3 \Gamma \left( \frac{3}{2} + s \right) }{\left( s+ \frac{1}{2} \right)^3\Gamma \left( 1-s \right)} (Ct)^s \right) \, ,
\end{align*}
where in the last line we use the defining identity of the gamma function as in \S \ref{return} to write $\Gamma \left( -\frac{1}{2} -s \right) = \frac{\Gamma \left( \frac{1}{2} - s \right)}{- \frac{1}{2} - s}$.  Since the function $f_1(s) \equiv \frac{\left( \Gamma \left( \frac{1}{2} -s \right) \right)^3 \Gamma \left( \frac{3}{2} + s \right)}{\Gamma\left( 1-s \right)}$ is analytic at $s=-\frac{1}{2}$, as is $(Ct)^s$ provided a branch is specified for $\log(\cdot)$, the residue at $s=-\frac{1}{2}$ is given by
\begin{align*}
\mathop{\mathrm{Res}}_{s=-\frac{1}{2}} \frac{f_1(s) (Ct)^s}{\left( s+\frac{1}{2} \right)^3} &= \left. \frac{1}{2!} \frac{d^2}{ds^2} \left[ f_1(s) (Ct)^s \right] \right|_{s = -\frac{1}{2}} \\
&= \frac{(Ct)^{-\frac{1}{2}}}{2} \left( f_1''\left(-\frac{1}{2}\right) + 2 f_1'\left( - \frac{1}{2} \right) \log(Ct) + f_1 \left( - \frac{1}{2} \right) (\log(Ct))^2 \right) \, ,
\end{align*}
where the values $f_1( -\frac{1}{2})$, $f_1'( -\frac{1}{2})$, $f_1''( -\frac{1}{2})$ are readily verified to be nonzero.  Thus, combining with the prefactor of $Ct$ as in \eqref{firsttermnu=0}, we see that the asymptotic behavior of the first term in the expression \eqref{MSD_int} for the mean squared displacement is of the order $t^{\frac{1}{2}}(\log(t))^2$.

We now expand the Meijer G-function $ \MeijerG*{3}{1}{2}{3}{1 , \frac{3}{2}}{1 , 1, 1}{ \rho t } $ appearing as an integrand in the $k \ne 0$ terms of the expression \eqref{MSD_int} for $\nu=0$ (as before, $\rho = \frac{16 \Lambda \pi^2}{\hbar^2} (\omega +2\pi k)^{-2}$).  Applying the residue theorem in a fashion similar to the above calculation for the $k=0$ term, we obtain
\begin{align*}
    \MeijerG*{3}{1}{2}{3}{1 , \frac{3}{2}}{1 , 1, 1}{ \rho t } &= - \sum_{a=1}^\infty \mathop{\mathrm{Res}}_{s=a} \left( \frac{\left( \Gamma \left(1-s\right) \right)^3 \Gamma(s) }{\Gamma \left( \frac{3}{2} - s \right)} \left( \rho t \right)^s \right) \\
    &= \frac{\rho t }{2} \left( g_1''(1) + 2 g_1'(1)  \log \left(\rho t \right) + g_1(1) \left(\log(\rho t )\right)^2 \right) + \\
    & \quad \frac{\left( \rho t \right)^2}{2} \left( g_2''(2) + 2 g_2'(2) \log \left( \rho t \right) + g_2(2) \left( \log \left( \rho t \right) \right)^2 \right) + \cdots \, , 
\end{align*}
where as in \S \ref{return}, $g_1(s) = \frac{\left( 
\Gamma (2-s) \right)^3 \Gamma \left( s \right)}{\Gamma \left( \frac{3}{2}-s \right)}$, $g_2(s) = \frac{\left( 
\Gamma (3-s) \right)^3 \Gamma \left( s \right)}{\left( 1-s \right)^3\Gamma \left( \frac{3}{2}-s \right)}$, and subsequent $g_n(s)$ follow from further applications of the defining relation of the gamma function to enable computation of residues.  Writing $\log(\rho t) = \log(\rho) + \log(t)$, it is evident that all dependence of the Meijer G-function on $t$ can be factored out of the integration in \eqref{MSD_int}, so that we can identify the dominant $t$-dependence of $ \langle \Delta \omega ^2 \rangle $ for small $t$ as $t^{\frac{1}{2}} \left( \log(t) \right)^2$ when $\nu= 0$.

\section{Discussion and conclusions}
\label{discussion}

\subsection{Anomalous diffusion and dimension}
In the preceding sections, we found that the value of the factor ordering parameter $\nu$ affects short-time properties of diffusion according to the quantum heat kernel.  We can now use these results to characterize quantized space at short distance scales, since diffusion over short times probes space over short distances. 

\begin{figure}[ht]
    \centering
    \includegraphics[width=0.7\textwidth]{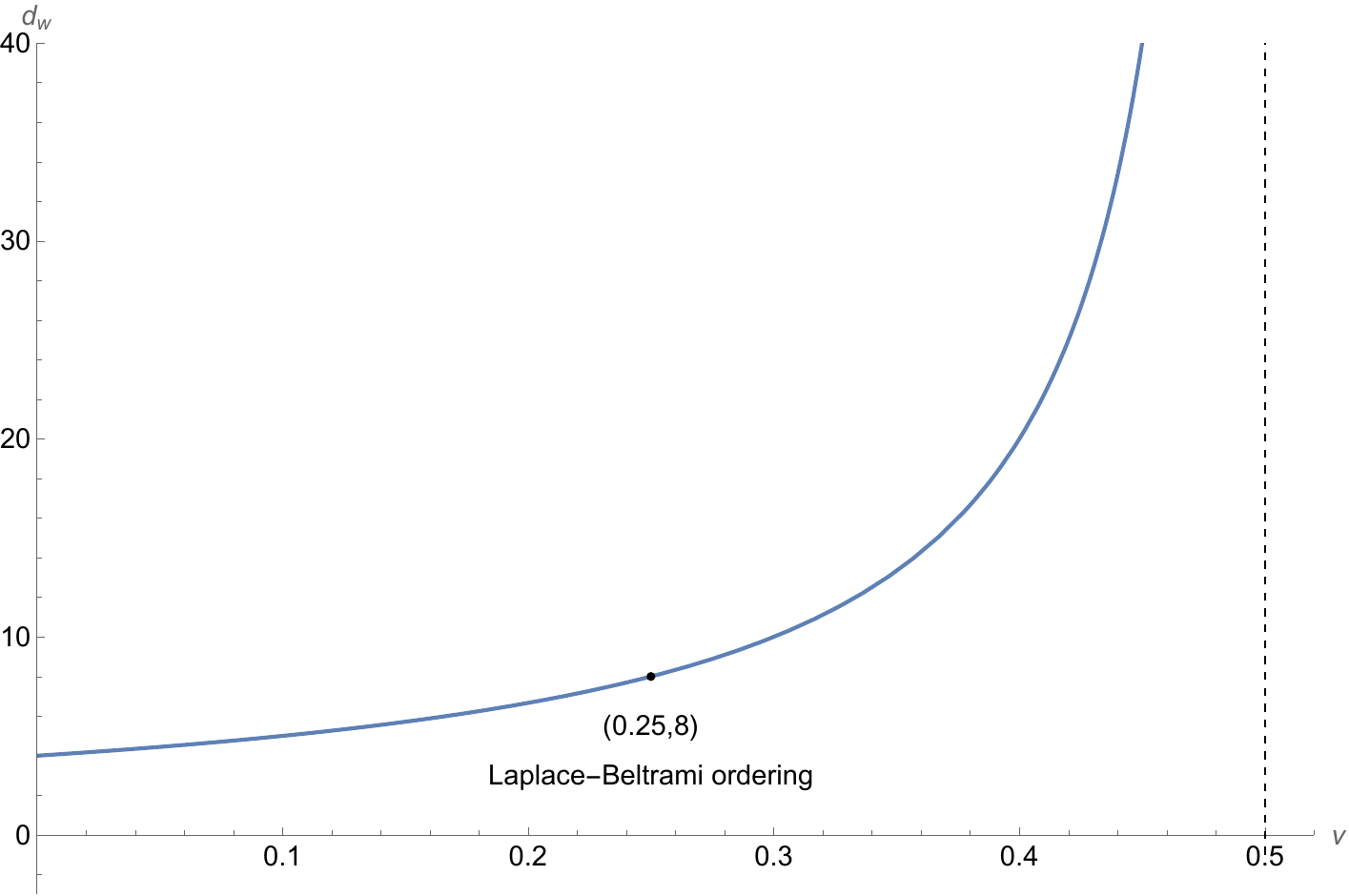}
    \caption{Walk dimension for the path of a diffusing particle as a function of ordering parameter $\nu$.  Note that the Laplace-Beltrami ordering ($\nu=\frac{1}{4}$) corresponds to $d_w=8$, while orderings close to the symmetric ordering ($\nu=0$) yield values of $d_w$ approaching 4.}
    \label{walkdimfig}
\end{figure}

The subdiffusive behavior evidenced by mean squared displacement \eqref{diffalpha} for $\nu \in ( 0, \frac{1}{2} )$ can be used to compute the walk dimension for the path of a diffusing particle according to \eqref{walkdim}:
\begin{align*}
&d_w = \frac{2}{\alpha} \, , \quad \alpha = \frac{1}{2} - \nu \\
\implies &d_w = \frac{4}{1-2 \nu}
\end{align*}
(see Figure \ref{walkdimfig}).  Note that because $\langle \Delta \omega^2 \rangle$ does not behave asymptotically as a power of $t$ for $\nu=0$, the same analysis cannot extend to that case.
 
By way of interpretation, note first that the infimum of 4 for $d_w$ is twice the standard walk dimension of 2 for non-anomalous ($\alpha=1$) diffusion, indicating that factor ordering ambiguity aside, quantization has resulted in more convoluted trajectories for particles diffusing over short times.  Second, as observed in our analysis of the return probability expansion at the end of \S \ref{heatkernel}, the Big-Bang universe wavefunction $\tilde \Psi_0 (\ell)$ increases more sharply as $\ell$ approaches 0 for factor orderings with larger values of $\nu$. 
Thus the larger the value of $\nu$, the likelier a diffusing particle is to find itself in tiny universes and to become ever more enmeshed in what is effectively at short distance scales a foamy, non-manifoldlike space.  The fact that the walk dimension approaches infinity as $\nu \to \frac{1}{2}$ mirrors the fact that for $\nu \ge \frac{1}{2}$, the wavefunction for a Big-Bang universe as defined in \eqref{BBwavefunctionsym} fails to be normalizable due to its asymptotic behavior near $\ell=0$.  

Using the exponent $\alpha = \frac{1}{2}-\nu$ for anomalous diffusion and the exponent $\beta = -\frac{1}{2}$ for the asymptotic behavior $P_0(t) \propto t^\beta$ of return probabilities for small time, we can similarly compute the short-distance effective dimension $\bar d$ of quantized space mentioned in \S\ref{dim} as mimicking Hausdorff dimension:
\begin{equation*}
\bar d = - \frac{2 \beta}{\alpha} = \frac{2}{1-2\nu} \, .
\end{equation*} 
In keeping with our observations on the walk dimension for a diffusing particle, we see that $\bar d$ has an infimum of 2, twice the classical spatial dimension prior to quantization.  A choice of ordering parameter $\nu$ sufficiently close to $\frac{1}{2}$ makes $\bar d$ arbitrarily large.  It is interesting to note that the Laplace-Beltrami ordering ($\nu=\frac{1}{4}$) yields integer values of 8 for the walk dimension $d_w$ and 4 for the effective dimension $\bar d$, while for orderings close to the symmetric ordering ($\nu=0$), $d_w$ approaches 4 and $\bar d$ approaches 2.

\subsection{Quantum geometry of a Big-Bang universe in (1+1) dimensions}
The preceding analysis shows that as the factor ordering parameter $\nu$ dictates an increased bias of our Big-Bang universe wavefunction toward small universes, diffusion becomes increasingly anomalous and the walk dimension of the path of a diffusing particle accordingly grows.  We are led to infer that the quantized space described by our wavefunction behaves less and less like a manifold as $\nu$ increases.  

While direct comparison with other investigations of quantized 2-dimensional spacetime is complicated by the delicacy of the various definitions for dimension, we can highlight qualitative parallels.  First, observe that as noted earlier, the term $\frac{\hbar^2 J_+^2}{4} \ell^{-1}$ arising from factor ordering in our Hamiltonian \eqref{QHamsym} is of the same form as the $\nu^2$ term in the minisuperspace Wheeler-DeWitt equation (3.13)
\begin{equation*}
\left[ - \left( \ell \frac{\partial}{\partial \ell} \right)^2 + 4 \mu \ell^2 + \nu^2 \right] \psi = 0 \, 
\end{equation*}
given in \citet{moore_loops_1991} (see also \citet{seiberg_notes_1990}, equation 4.9).  This term depends upon the parameter $Q$ introduced through renormalization and appearing in the quantum version of the Liouville action as a coefficient on the curvature $R(\hat g)$ of the reference metric $\hat g$ \citep{seiberg_notes_1990}.  In principle, given a cylindrical spacetime topology, the reference metric could have been chosen to be flat and the $R(\hat g)$ term eliminated from the action at the classical level.
That the factor ordering ambiguity independently introduces an analogous term invites speculation as to whether choice of factor ordering can be viewed as tuning the degree of quantum-level disruptions to spatial topology.  The change to walk dimension of the path of a particle diffusing according to our quantum heat kernel can be viewed as the manifestation of such disruptions.

Computations in causal dynamical triangulations for 2-dimensional spacetime demonstrate that allowing specific types of topological disruption indeed increases the Hausdorff dimension of quantized spacetime.  \citet{ambjorn_non-perturbative_1998} show that topology change and proliferation of baby universes affect the Hausdorff dimension of quantized 2-dimensional spacetime by raising it from 2 to 4.  %While the continuum expression for the quantized Hamiltonian derived in \citet{ambjorn_non-perturbative_1998} is the same as our \eqref{QHamiltonian} with ordering parameter $J_+=1$ (equivalently $\nu = \frac{1}{2}$) and thus lies outside the domain of applicability of our results since there is no normalizable solution to the Wheeler-DeWitt equation for $J_+ \ge 1$, we note the general conclusion that topology change increases Hausdorff dimension.  
Additionally, numerical simulations in \citet{loll_locally_2015} suggest that relaxing the requirement of a preferred global foliation similarly increases the computed Hausdorff dimension.

As well as analyzing the anomalous behavior of diffusion and associated short-distance properties of space as quantum effects, it is well to reflect on the role played by the particular spatial topology $S^1$ of our classical model.  The heat kernel \eqref{heat} on $S^1$ is written as an infinite series, with the index $k$ referring to the number of complete circuits made by the diffusing particle.  This structure persists through the calculations of the quantum heat kernel \eqref{qheat}, and subsequently the return probability \eqref{Ptbeta} and mean squared displacement \eqref{MSD_int}.  The effect of factor ordering on return probability is twofold:  the effective diffusion coefficient read off from the first term depends on $\nu$, and the higher-order terms in the expansion resulting from diffusion around more than one complete circuit are modified from a power series form.  While the first effect is not directly connected to the spatial topology of the universe, the second clearly is, indicating that the non-manifold-like behavior manifested in the return probability arises from the ability of a diffusing particle to complete multiple circuits around the spatial $S^1$.  The effect of factor ordering on mean squared displacement, on the other hand, arises from the asymptotics of the individual terms in the heat kernel, and thus is not directly connected to the fact that a diffusing particle can make multiple circuits around the universe.

In seeking to understand the origins of the observed effects of quantization on diffusion, we must also bear in mind that ours is a minisuperspace analysis in the sense that the quantum heat kernel is constructed from heat kernels for diffusion on spaces of strictly circular geometry.  On the other hand, the action \eqref{Lagrangian} is shown in \citet{nakayama_2d_1994} to arise not simply from reduction to a minisuperspace model by the imposition of circular spatial symmetry, but as a reduced model obtained from the full Hamiltonian by fixing a proper-time gauge and integrating out the spatial dependence of the space-space metric component.  From this perspective, we can regard the heat kernel for a circle of circumference $\ell$ as representing the equivalence class of heat kernels on spatial manifolds with topology $S^1$ and arclength $\ell$.  However, this argument is heuristic, and ultimately our expected heat kernel must be regarded as a model, yielding precise results in the minisuperspace case with circular symmetry imposed classically on spatial slices. 

\subsection{Future work:  Expected heat kernel for quantum cosmologies}

Toward generalizing the results of the present paper, we note that the non-Gaussian anomalous diffusion we observe is tied to the topology of the spatial universe $S^1$.  However, in any universe with a Big Bang (or Big Crunch), the distortion of spacetime close to these extreme events must be expected to affect diffusion; for example in \citet{smerlak_diffusion_2012} and \citet{bonifacio_brownian_2012}, diffusion is studied on general curved backgrounds as well as specifically in a Friedmann-Lema\^{i}tre-Robertson-Walker (FLRW) cosmology with negative, positive, or zero curvature.  In turn, a wavefunction for a universe propagating from a Big Bang scenario (or toward a Big Crunch) will include nonzero weights for spacetimes arbitrarily close to the singularity, with factor ordering expected to play a role in the asymptotic behavior of the wavefunction near the singularity owing to the connection of such asymptotics with membership in a self-adjoint extension of the Wheeler-DeWitt operator. 

To sketch this anticipated dependence, we consider the case of FLRW quantum cosmology with positive curvature.  For the spacetime metric on $\R \times S^3$ in the ADM formalism with scale factor $a(t)$, given by
\begin{equation*}
ds^2 = -N^2(t) \, dt^2 + a^2(t) \left( d \theta_1^2 + \sin^2(\theta_1) \left( d \theta_2^2 + \sin^2(\theta_2) d \varphi^2 \right) \right) \, ,
\end{equation*}
where $N(t)$ represents the lapse function, the Hamiltonian constraint $\mathcal{H} = G_{ijkl} \pi^{ij} \pi^{kl} - h^{1/2} \left( ^3R -2\Lambda \right) = 0$ leads to the effective action
\begin{align*}
L = \frac{1}{2} \left( -a \dot a^2 + a - \frac{\Lambda}{3} a^3 \right)
\end{align*}
with the choice of lapse $N=1$, and associated Hamiltonian 
\begin{equation*}
H = \frac{1}{2} \left( -\frac{\Pi_a^2}{a} - a + \frac{\Lambda}{3} a^3  \right) \, ,
\end{equation*}
where $\Pi_a = -a \dot a$.  Under a Schr\"{o}dinger ansatz $\hat a \psi = a \psi(a)$, $\hat \Pi_a \psi = - i \hbar \frac{d \psi}{d a}$, we can allow for a range of factor orderings analogous to that we have considered for the case of $(1+1)$-dimensional quantum gravity:
\begin{align*}
\hat{H} &= \frac{1}{2} \left(- a^{j_1} \hat \Pi_a a^{j_2} \hat \Pi_a a^{j_3} -a + \frac{\Lambda}{3} a^3 \ \right) \, , \quad j_1 + j_2 + j_3 = -1 \notag \\
&= \frac{1}{2} \left(  \hbar^2 a^{j_1} \frac{d}{da} a^{-1- \left( j_1 + j_3 \right)} \frac{d}{da} a^{j_3} - a + \frac{\Lambda}{3} a^3 \right) \, .
\end{align*}
As for (1+1)-dimensional quantum gravity, we can define $\tilde H = a^{J_-/2} \hat H a^{-J_-/2}$ symmetric with respect to the Lebesgue measure, where $J_\pm \equiv j_3 \pm j_1$.  The Wheeler-DeWitt equation $\tilde H \psi = 0$ then yields 
\begin{equation}
\hbar^2 \left( a^2\psi'' - a\psi' - \frac{J_+}{2} \left( 2 + \frac{J_+}{2} \right) \psi \right) + \left( -a^4 + \frac{\Lambda}{3} a^6 \right) \psi = 0 \, ,
\label{WDW-FLRW}
\end{equation}
in which for small $a$ the term $\left( -a^4 + \frac{\Lambda}{3} a^6 \right) \psi$ has dominant behavior given by $-a^4 \psi$.  Neglecting $\frac{\Lambda}{3} a^6 \psi$ and carrying out the transformation $\psi(a) = z^{1/2} \phi(z)$, where $z= \frac{a^2}{2 \hbar}$, yields a modified Bessel equation with index $\nu = \frac{1}{2}\sqrt{1 + \frac{J_+}{2} \left( 2 + \frac{J_+}{2} \right)}$, so that asymptotically for small $a$, solutions to the Wheeler-DeWitt equation \eqref{WDW-FLRW} are given by the modified Bessel solution 
\begin{equation*}
\psi \sim C_1 I_\nu \left( \frac{a^2}{2 \hbar} \right) + C_2 K_\nu \left( \frac{a^2}{2 \hbar} \right) \, .
\end{equation*}
It is clear that as in the case of (1+1)-dimensional quantum gravity, the asymptotic behavior of wavefunctions for small universe depends on the factor ordering parameter $J_+$.  Solutions to the Wheeler-DeWitt equation for the FLRW cosmology with an equivalent range of orderings have been studied in \citet{kontoleon_operator_1999}, \citet{wiltshire_wave_2000}, and \citet{steigl_factor_2006}, similarly finding dependence on factor ordering of wavefunction asymptotics for $a \to 0$.

In the case of (1+1)-dimensional quantum gravity, the dependence on $J_+$ of the asymptotic behavior of the wavefunction for propagation from a Big Bang led to anomalous diffusion for the quantum heat kernel and a small-scale walk dimension determined by choice of factor ordering.  For FLRW cosmology, the analogous behavior of solutions to the Wheeler-DeWitt equation indicates the possibility of similar consequences for diffusion according to a quantum heat kernel.  The heat kernel expansion on a sphere of arbitrary dimension is known \citep{zhao_exact_2018, kluth_heat_2020}, and while computation of the quantum heat kernel for the FLRW cosmology would likely necessitate a more numerical approach than in the present paper, investigating its behavior is an intriguing prospective focus for future research.  

Finally, we note the ubiquity of the factor ordering problem for quantum gravity.  Not only does this question attend any instance of the Wheeler-DeWitt equation, whether for quantum cosmology or full quantum gravity, but a path integral formulation also offers no escape, since the dual ambiguity is in the definition of the path integral measure.  By studying the consequences of the choice of factor ordering for diffusion behavior according to an expected heat kernel, we hope to uncover clues which might guide the choice using physical consequences for quantum geometry.  

%For the spacetime metric on $\R \times S^3$ in the ADM formalism with scale factor $a(t)$, given by
%\begin{equation*}
%ds^2 = -N^2(t) \, dt^2 + a^2(t) \left( d \theta_1^2 + \sin^2(\theta_1) \left( d \theta_2^2 + \sin^2(\theta_2) d \varphi^2 \right) \right) \, ,
%\end{equation*}
%where $N(t)$ represents the lapse function, the Hamiltonian constraint $\mathcal{H} = G_{ijkl} \pi^{ij} \pi^{kl} - h^{1/2} \left( ^3R -2\Lambda \right) = 0$ leads to the effective action
%\begin{align*}
%L = \frac{1}{2} \left( -a \dot a^2 + a - \frac{\Lambda}{3} a^3 \right)
%\end{align*}
%with the choice $N=1$.  

%As in the present case, the weighting of the corresponding heat kernels may similarly lead to a quantum heat kernel exhibiting anomalous diffusion dependent on choice of factor ordering.  Investigating this possibility is an intriguing prospective focus for future research.

\section*{Acknowledgements}

The authors extend many thanks to Harold Erbin for helpful correspondence and for furnishing useful notes on 2d quantum gravity.  Additionally, the authors are grateful to the editor and anonymous reviewers for valuable feedback and for bringing a crucial reference to our attention.  Last but not least, we wish to thank Kelsey Diemand, Associate Director for Research and Collections at the Douglas D. Schumann Library and Learning Commons, Wentworth Institute of Technology, for expert reference help.

\appendix
\section{Special functions}

For convenience, we collect in this Appendix a number of well-known results pertaining to special functions used in the body of the paper.  For a comprehensive reference on special functions, the reader is referred e.g. to \citet{luke_special_1969}.

\subsection{Bessel functions}
Bessel functions are canonical solutions $y(z)$ of Bessel's differential equation
\begin{equation}
    z^2 \frac{d^2y}{dz^2} + z\frac{dy}{dz} + (z^2 - \nu^2)y = 0
    \label{besselDiff}
\end{equation}
for an arbitrary complex number $\nu$. As this is a second-order linear differential equation, there are two linearly independent solutions; however, we only consider the first of the two in our analysis. Bessel functions of the first kind are denoted by $J_\nu(z)$. If $\nu$ is an integer or positive, $J_\nu(z)$ is finite at the origin, while if $\nu$ is a negative non-integer then $J_\nu(z)$ diverges as $z \to 0$. We can define the Bessel function by a series expansion about $z=0$ which is obtained from applying the Frobenius method to Bessel's differential equation:
\begin{equation*}
    J_\nu(z) = \sum_{m=0}^\infty \frac{(-1)^m}{m! \Gamma(m+\nu+1)} \left(\frac{z}{2}\right)^{2m+\nu}.
\end{equation*}
Since the Bessel functions are defined for complex argument $z$, a special case arises when $z$ is purely imaginary. Here, the solutions to \eqref{besselDiff} are called the modified Bessel functions of the first and second kind and are defined as
\begin{align*}
    I_\nu(z) &= i^{-\nu} J_\nu(iz) = \sum_{m=0}^\infty \frac{1}{m!\Gamma(m+\nu+1)} \left(\frac{z}{2}\right)^{2m+\nu}, \\
    K_\nu(z) &= \frac{\pi}{2}\frac{I_{-\nu}(z) - I_\nu(z)}{\sin(\nu z)},
\end{align*}
when $\nu$ is not an integer; otherwise, the limit is used in the definition of $K_\nu$.  The modified Bessel functions satisfy the modified Bessel's equation $z^2 \frac{d^2y}{dz^2} + z \frac{dy}{dz} - (z^2 + \nu^2) y = 0$.

Note that the asymptotic behavior of the Bessel functions as $z \to 0$ follows directly from their series definition for $\nu \ne 0$, and for $\nu=0$ from the limiting definition.   For use in the present paper, we note in particular 
\begin{equation}
\begin{split}
I_\nu (z) &\sim \frac{1}{\Gamma(\nu+1)} \left( \frac{1}{2} z \right)^\nu \, , \quad \nu \ne -1,-2, -3, \dots\\
K_\nu (z) &\sim \frac{1}{2} \Gamma( \lvert \nu \rvert ) \left( \frac{1}{2} z \right)^{- \lvert \nu \rvert} \, , \quad  \nu \ne 0 \\
K_0(z) &\sim -\log(z)
\label{Besselasymp}
\end{split}
\end{equation}

% One can represent the Bessel function of the first kind in terms of the modified Bessel functions, specifically if $-\pi < \arg z \leq \frac{\pi}{2}$ then
% \begin{equation}
%     J_\alpha(iz) = e^{\frac{\alpha\pi i}{2}} I_\alpha(z). \label{ItoJ}
% \end{equation}

\subsection{Generalized hypergeometric functions}
A generalized hypergeometric function is defined by a convergent generalized hypergeometric series
\begin{equation}
    \pFq{p}{q}{a_1,\dots,a_p}{b_1,\dots,b_q}{z} = \sum_{k=0}^\infty \frac{(a_1)_k\cdots(a_p)_k}{(b_1)_k\cdots(b_q)_k}\frac{z^k}{k!},
\label{pFq}
\end{equation}
where $(x)_k$ is the Pochhammer symbol or rising factorial $(x)_k = x(x+1)(x+2)\cdots(x+k+1)$. Of course, depending on the choice of $a_i$ and $b_j$, \eqref{pFq} could yield a terminating or undefined series (namely, if some $a_i$ is a non-positive integer we have the former and if some $b_j$ is a non-positive integer we have the latter). The radius of convergence is dependent on the choice of $p,q$, by use of the ratio test. If $p < q+1$ then the ratio of coefficients approaches zero, which immediately gives convergence for any finite $z$ and hence defines an entire function. If $p = q+1$ then the ratio of coefficients tends to one, so the series converges for $|z| < 1$ and diverges for $|z| > 1$. Convergence for $|z| = 1$ is more difficult to determine. Finally, if $p > q+1$, then the ratio of coefficients grows toward infinity. Thus, away from $z = 0$, the series diverges. 

A wide variety of elementary and special functions can be expressed in terms of generalized hypergeometric functions; for our purposes, a relevant example is the following expression of the square of a Bessel function as a generalized hypergeometric function, as shown here from \citet{luke_special_1969} \S6.2:
\begin{align}
    J_\nu^2 (z) = \frac{(\frac{z}{2})^{2\nu}}{[\Gamma(\nu + 1)]^2} \cdot \pFq{1}{2}{\nu + \frac{1}{2}}{\nu + 1, 2 \nu + 1}{-z^2}.
    \label{BesselHyper}
\end{align}

A strong advantage of working in terms of generalized hypergeometric functions is the fact that basic properties such as derivatives and antiderivatives follow immediately from the definition \eqref{pFq}, specifically
\begin{equation}
    \int z^{\alpha - 1} \, \pFq{p}{q}{a_1,\dots,a_p}{b_1,\dots,b_q}{z} \, dz = \frac{z^\alpha}{\alpha} \, \pFq{p+1}{q+1}{\alpha,a_1,\dots,a_p}{\alpha+1,b_1,\dots,b_q}{z}.
    \label{pFqInt}
\end{equation}

\subsection{Meijer G-functions}
Meijer G-functions are defined in terms of a contour integral as
\begin{equation}
\MeijerG*{m}{n}{p}{q}{a_1 , \dots , a_p}{b_1 , \dots , b_q}{z} = {\frac {1}{2\pi i}}\int _{L}{\frac {\prod _{j=1}^{m}\Gamma (b_{j}-s)\prod _{j=1}^{n}\Gamma (1-a_{j}+s)}{\prod _{j=m+1}^{q}\Gamma (1-b_{j}+s)\prod _{j=n+1}^{p}\Gamma (a_{j}-s)}}\,z^{s}\,ds,
\label{MeijerGDef}
\end{equation}
where the contour $L$ is chosen in a canonical fashion based on the values of $p$, $q$, $m$, and $n$ so as to make the integral converge (for details, see e.g. \citet{luke_special_1969} \S5.2).  For our purposes, the relevant case will be that where $q \ge 1$ and $p<q$.  Under these conditions, $L$ is taken to be the loop beginning and ending at $+\infty$ and encircling all poles of $\Gamma(b_j - s)$, $j=1, \dots ,m$, exactly once in the negative direction, but not encircling any pole of $\Gamma(1-a_k+s)$, $k=1, \dots , n$.  (The same contour also applies in the case where $q \ge 1$, $p=q$, and $|z|<1$.)

Most commonly used elementary and special functions can be represented as Meijer G-functions.  Identifications used in the present paper include the following relationship between a product of modified Bessel functions of the second kind and a Meijer G-function:
\begin{equation}
z^\omega K_\mu (z) K_\nu(z) = \frac{\sqrt{\pi}}{2} \MeijerG*{4}{0}{2}{4}{\frac{\omega}{2} , \frac{\omega+1}{2}}{\frac{\omega + \mu + \nu}{2}, \frac{\omega - \mu + \nu}{2}, \frac{\omega + \mu - \nu}{2}, \frac{\omega - \mu - \nu}{2} }{z^2}
\label{KtoG}
\end{equation}
(see e.g. \citet{luke_special_1969} \S 6.4 (33)).

Basic identities for the Meijer G-function follow straightforwardly from the definition \eqref{MeijerGDef}. In particular, if we have equality of two parameters $a_k$, $k=1, \dots , n$ and $b_j$, $j=m+1 , \dots , q$, the corresponding factors cancel and the order of the Meijer G-function is lowered:
\begin{equation*}
\MeijerG*{m}{n}{p}{q}{a_1 , \dots , a_k, \dots , a_p}{b_1 , \dots , b_j , \dots , b_q}{z} = \MeijerG*{m}{n-1}{p-1}{q-1}{a_1 , \dots , a_{k-1}, a_{k+1}, \dots , a_p}{b_1 , \dots , b_{j-1} , b_{j+1} , \dots , b_q}{z}, \quad n,p,q \ge 1 \, ;
\end{equation*}
similarly, if equality holds between parameters $a_k$, $k=n+1, \dots , p$, and $b_j$, $j=1, \dots , m$, then 
\begin{equation}
\MeijerG*{m}{n}{p}{q}{a_1 , \dots , a_k, \dots , a_p}{b_1 , \dots , b_j , \dots , b_q}{z} = \MeijerG*{m-1}{n}{p-1}{q-1}{a_1 , \dots , a_{k-1}, a_{k+1}, \dots , a_p}{b_1 , \dots , b_{j-1} , b_{j+1} , \dots , b_q}{z}, \quad m,p,q \ge 1 \, . \label{GReduction}
\end{equation}
The substitution $s+\sigma \to s$ in the defining contour integral yields the identity 
\begin{equation}
z^\sigma \MeijerG*{m}{n}{p}{q}{a_1 , \dots , a_p}{b_1 , \dots , b_q}{z} = \MeijerG*{m}{n}{p}{q}{a_1 + \sigma , \dots , a_p+\sigma}{b_1+\sigma , \dots , b_q+\sigma}{z} \, .
\label{MeijerPowerProd}
\end{equation}

Under the condition $p<q$, or alternatively $p=q$ and $|z|<1$, the contour integral \eqref{MeijerGDef} can be evaluated using the Cauchy residue formula to obtain an expression for the Meijer G-function in terms of generalized hypergeometric functions.  If no two of the $b_j$, $j=1,\dots,m$ are equal or differ by an integer, the poles are all simple, leading to the result
\begin{align}
\begin{split}
\MeijerG*{m}{n}{p}{q}{a_1 , \dots , a_p}{b_1 , \dots , b_q}{z} = \sum_{h=1}^m \frac{\prod_{j=1}^m \Gamma (b_j - b_h)^\ast \prod_{j=1}^n \Gamma(1+b_h-a_j) z^{b_h}}{\prod_{j=m+1}^q \Gamma (1+b_h-b_j) \prod_{j=n+1}^p \Gamma(a_j-b_h)} \, \times \\  \pFq{p}{q-1}{1+b_h-a_j}{1+b_h-b_q \,^*}{(-1)^{p-m-n}z} \, .
\end{split}
\label{MeijerG_F}
\end{align}
Note that as a consequence, for $p \le q$,
\begin{equation*}
\MeijerG*{m}{n}{p}{q}{a_1, \dots , a_p}{b_1 , \dots , b_q}{z} = O(\lvert z \rvert ^\beta) \, , \quad \beta = \min \Re (b_1 , \dots , b_q )
\end{equation*}
for $z$ close to 0, since the requirement $p \le q$ ensures generalized hypergeometric functions $_pF_{q-1}$ are defined by power series expansions convergent for $|z|<1$ (see \eqref{pFq}).

By using the definition \eqref{MeijerGDef} along with the functional relation for the gamma function $z \Gamma (z) = \Gamma(z+1)$, one can compute derivatives and antiderivatives of various products of Meijer G-functions with powers, for example
\begin{equation}
\int z^{-a_p-1} \MeijerG*{m}{n}{p}{q}{a_1 , \dots , a_p}{b_1 , \dots , b_q}{z} \, dz = -z^{-a_p} \MeijerG*{m}{n}{p}{q}{a_1 , \dots , a_{p-1} , a_p+1}{b_1 , \dots , b_q}{z} \, ,
\label{MeijerIndefInt}
\end{equation}
valid for $n<p$ (see e.g. \citet{prudnikov_integrals_1986} \S1.16.2 (3)).  Identities for the Meijer G-function deriving from its definition lead to the related equality 
\begin{equation}
\int_0^z y^{s-1} \MeijerG*{m}{n}{p}{q}{a_1 , \dots , a_p}{b_1 , \dots , b_q}{\eta y} \, dy = z^s \MeijerG*{m}{n+1}{p+1}{q+1}{1-s , a_1 , \dots , a_p}{b_1 , \dots , b_q, -s}{\eta z} \, ,
\label{MeijerInt0toz}
\end{equation}
valid under a variety of conditions, the relevant one for our purposes being
\begin{align*}
    a_l - b_k \not\in \N \ (k = 1,\dots,m, \ l = 1,\dots,n) \, ; \quad \Re (s) + \min_{1 \le j \le m} \Re (b_j) > 0 \,; \\
    \quad \delta , z > 0 \, ; \quad |\arg\eta| < \delta\pi \, ,
\end{align*}
where $\delta \equiv m+n-\frac{p+q}{2}$ (see e.g. \citet{prudnikov_integrals_1986} \S1.16.2 (1)).

By relating the integral in the definition \eqref{MeijerGDef} to an inverse Mellin transform, one can use the Mellin inversion theorem to evaluate the definite integral of a power times the Meijer G-function
\begin{equation}
\int_0^\infty z^{s-1} \MeijerG*{m}{n}{p}{q}{a_1 , \dots , a_p}{b_1 , \dots , b_q}{\eta z} \, dz = \frac{\eta^{-s} \prod_{j=1}^m \Gamma (b_j+s) \prod_{j=1}^n \Gamma (1-a_j-s) }{ \prod_{j=m+1}^q \Gamma (1-b_j-s) \prod_{j=n+1}^p \Gamma (a_j+s) } \, ,
\label{MeijerDefInt}
\end{equation}
valid in a variety of cases outlined for example in \citet{luke_special_1969}.  For our purposes, the relevant case is as follows:
\begin{align}
\begin{split}
    n = 0; \quad 1 \leq p + 1 \leq m \leq q; \quad -\min_{1 \leq h \leq m} \Re(b_h) < \Re(s) < 1 - \max_{1\leq j \leq n} \Re(a_j); \\
    \delta > 0; \quad |\arg\eta| < \delta\pi.
    \label{GIntCase2}
\end{split}
\end{align}

\section{Normalizing the wavefunction}
\label{appendix-normalization}
We outline in this Appendix the technical calculation of the integral in \eqref{I1I2}. Denote the first integral in the last line of \eqref{I1I2} together with its Bessel function coefficient as $\mathcal{I}_1$. This integral is easily calculated by rewriting it as a Bessel function of the first kind, which in turn allows conversion to a generalized hypergeometric function thanks to \eqref{BesselHyper}. Some simplification yields
\begin{align*}
    \mathcal{I}_1 &= K^2_{\nu}\left(\tfrac{\sqrt{\Lambda}}{\hbar}\ell_*\right) \int_0^{\frac{\sqrt\Lambda \ell_*}{\hbar}} I_\nu^2(u) \, du \\
    &= i^{-2\nu} K^2_{\nu}\left(\tfrac{\sqrt{\Lambda}}{\hbar}\ell_*\right) \int_0^{\frac{\sqrt\Lambda \ell_*}{\hbar}} J_\nu^2(iu) \, du \\
    &= \frac{K^2_{\nu}\left(\tfrac{\sqrt{\Lambda}}{\hbar}\ell_*\right)}{2^{2\nu}[\Gamma(\nu+1)]^2} \int_0^{\frac{\sqrt{\Lambda}\ell_*}{\hbar}} u^{2\nu} \pFq{1}{2}{\nu+\frac{1}{2}}{\nu+1, 2\nu+1}{u^2} \, du.
\end{align*}
Making another substitution $z = u^2$, applying \eqref{pFqInt} by setting $\alpha = \nu + \frac{1}{2}$, and simplifying yields
\begin{align*}
    \mathcal{I}_1 &= \frac{K^2_{\nu}\left(\tfrac{\sqrt{\Lambda}}{\hbar}\ell_*\right)}{2^{2\nu+1}[\Gamma(\nu+1)]^2} \int_0^{\frac{\Lambda\ell_*^2}{\hbar}} z^{\nu - \frac{1}{2}} \, \pFq{1}{2}{\nu+\frac{1}{2}}{\nu+1, 2\nu+1}{z} \, dz \\
    &= \frac{K^2_{\nu}\left(\tfrac{\sqrt{\Lambda}}{\hbar}\ell_*\right)}{2^{2\nu+1}[\Gamma(\nu+1)]^2} \, \frac{\left(\frac{\Lambda \ell_*^2}{\hbar^2}\right)^{\nu + \frac{1}{2}}}{\nu + \frac{1}{2}} \pFq{2}{3}{\nu + \frac{1}{2}, \nu + \frac{1}{2}}{\nu + \frac{3}{2}, \nu + 1, 2\nu + 1}{\frac{\Lambda \ell_*^2}{\hbar^2}}.
\end{align*}
We direct our attention to the second integral in \eqref{I1I2} and denote it $\mathcal{I}_2$, similarly including its Bessel function coefficient. Instead of converting to a generalized hypergeometric function, we will convert to a Meijer G-function via \eqref{KtoG}. This yields
\begin{align*}
    \mathcal{I}_2 &= I_\nu^2\left(\frac{\sqrt\Lambda}{\hbar}\ell_*\right) \int_{\frac{\sqrt\Lambda \ell_*}{\hbar}}^\infty K_\nu^2(u) \, du \\
    &= \frac{\sqrt\pi}{2} \, I_\nu^2\left(\frac{\sqrt\Lambda}{\hbar}\ell_*\right) \int_{\frac{\sqrt\Lambda \ell_*}{\hbar}}^\infty \, \MeijerG*{3}{0}{1}{3}{\frac{1}{2}}{\nu,0,-\nu}{u^2} \, du.
\end{align*}
Substituting $z = u^2$ and splitting the domain of integration yields
\begin{align}
\begin{split}
    \mathcal{I}_2 &= \frac{\sqrt\pi}{4} \, I_\nu^2\left(\frac{\sqrt\Lambda}{\hbar}\ell_*\right) \int_{\frac{\Lambda\ell_*^2}{\hbar}}^\infty z^{-\frac{1}{2}} \, \MeijerG*{3}{0}{1}{3}{\frac{1}{2}}{\nu,0,-\nu}{z} \, dz \\
    &= \frac{\sqrt\pi}{4} \, I_\nu^2\left(\frac{\sqrt\Lambda}{\hbar}\ell_*\right) \Bigg[ \int_0^\infty z^{-\frac{1}{2}} \, \MeijerG*{3}{0}{1}{3}{\frac{1}{2}}{\nu,0,-\nu}{z} \, dz - \\
    &\pushright{\int_0^{\frac{\Lambda\ell_*^2}{\hbar}} z^{-\frac{1}{2}} \, \MeijerG*{3}{0}{1}{3}{\frac{1}{2}}{\nu,0,-\nu}{z} \, dz \Bigg].}
    \label{I2split}
\end{split}
\end{align}
We first consider the improper integral with bounds zero and infinity. The equality (1) in \S5.6.1 from \citet{luke_special_1969} provides a direct evaluation of such integrals provided appropriate conditions are met; these conditions are outlined in the appendix of the present paper (see \eqref{MeijerDefInt}, \eqref{GIntCase2}). Specifically, we set $\eta = 1, \ s = \frac{1}{2}, \ m=3, \ n=0, \ p=1$, and $q=3$, so $\delta = 1$. The verification of such conditions for this case is straightforward. Thus, applying \eqref{MeijerDefInt} and using various properties of the gamma function yields
\begin{align*}
    \int_0^\infty z^{-\frac{1}{2}} \, \MeijerG*{3}{0}{1}{3}{\frac{1}{2}}{\nu,0,-\nu}{z} \, dz = \frac{\pi^{\frac{3}{2}}}{\cos(\pi\nu)}.
\end{align*}

For the remaining integral in \eqref{I2split}, the equality \eqref{MeijerInt0toz} provides yet again a direct evaluation of such integrals. Setting $s = \frac{1}{2}$, $(a_p) = (\frac{1}{2})$, $(b_p) = (\nu, 0, -\nu)$, $\eta=1$, $m=3$, $n=0$, $p=1$, and $q=3$, so $\delta = 1$, it is easy to verify that the required conditions are met. Thus, we obtain
\begin{align}
    \int_0^{\frac{\Lambda\ell_*^2}{\hbar}} z^{-\frac{1}{2}} \, \MeijerG*{3}{0}{1}{3}{\frac{1}{2}}{\nu,0,-\nu}{z} \, dz = \frac{\sqrt\Lambda \ell_*}{\hbar} \, \MeijerG*{3}{1}{2}{4}{\frac{1}{2},\frac{1}{2}}{\nu,0,-\nu,-\frac{1}{2}}{\frac{\Lambda \ell_*^2}{\hbar^2}} \, ,
    \label{termL1}
\end{align}
so that 
\begin{align*}
    \mathcal{I}_2 = \frac{\sqrt\pi}{4} \, I_\nu^2\left(\frac{\sqrt\Lambda}{\hbar}\ell_*\right) \left[\frac{\pi^{\frac{3}{2}}}{\cos(\pi\nu)} - \frac{\sqrt\Lambda \ell_*}{\hbar} \, \MeijerG*{3}{1}{2}{4}{\frac{1}{2},\frac{1}{2}}{\nu,0,-\nu,-\frac{1}{2}}{\frac{\Lambda \ell_*^2}{\hbar^2}}\right].
\end{align*}
Finally, 
\begin{align*}
\begin{split}
    \int_0^\infty \lvert \Psi_{\ell_*}^\pm(\ell) \rvert^2 \, \ell^{J_-} \, d \ell &= \frac{\ell_*^{-J_-}}{2\hbar\sqrt\Lambda} \Bigg[ \frac{\left(\frac{\Lambda \ell_*^2}{\hbar^2}\right)^{\nu+\frac{1}{2}}K_\nu^2\left(\frac{\sqrt\Lambda}{\hbar}\ell_*\right)}{2^{2\nu}[\Gamma(\nu+1)]^2(\nu + \frac{1}{2})} \, \pFq{2}{3}{\nu + \frac{1}{2}, \nu + \frac{1}{2}}{\nu + \frac{3}{2},\nu + 1,2\nu + 1}{\frac{\Lambda \ell_*^2}{\hbar^2}} + \\
    & \quad \frac{\sqrt\pi}{2} \, I_\nu^2\left(\frac{\sqrt\Lambda}{\hbar} \ell_*\right) \left( \frac{\pi^{\frac{3}{2}}}{\cos(\nu\pi)} - \frac{\sqrt\Lambda}{\hbar}\ell_* \, \MeijerG*{3}{1}{2}{4}{\frac{1}{2},\frac{1}{2}}{\nu,0,-\nu,-\frac{1}{2}}{\frac{\Lambda}{\hbar^2}\ell_*^2}\right) \Bigg] \\
    &\equiv  \mathcal{N}^{-2},
\end{split}
\end{align*}
and our normalized amplitude with respect to $\ell$ becomes $\mathcal N \Psi_{\ell_*}^\pm(\ell)$.

% \section*{References}
\bibliographystyle{elsarticle-harv}
\bibliography{newRef.bib}

\end{document}